# Power Allocation Games in Interference Relay Channels: Existence Analysis of Nash Equilibria

Elena Veronica Belmega, Brice Djeumou, and Samson Lasaulce

**Abstract**

We consider a network composed of two interfering point-to-point links where the two transmitters can exploit one common relay node to improve their individual transmission rate. Communications are assumed to be multi-band and transmitters are assumed to selfishly allocate their resources to optimize their individual transmission rate. The main objective of this paper is to show that this conflicting situation (modeled by a non-cooperative game) has some stable outcomes, namely Nash equilibria. This result is proved for three different types of relaying protocols: decode-and-forward, estimate-and-forward, and amplify-and-forward. We provide additional results on the problems of uniqueness, efficiency of the equilibrium, and convergence of a best-response based dynamics to the equilibrium. These issues are analyzed in a special case of the amplify-and-forward protocol and illustrated by simulations in general.

## I. Introduction

A possible way to improve the performance in terms of range, transmission rate or quality of a network composed of mutual interfering independent source-destination links, is to add some relaying nodes in the network. This approach can be relevant in both wired and wireless networks. For example, it can be desirable

E. V. Belmega, B. Djeumou, and S. Lasaulce are with LSS (joint lab of CNRS, Supélec, Paris 11), Supélec, Plateau du Moulon, 91192 Gif-sur-Yvette, France, {belmega,djeumou,lasaulce}@lss.supelec.fr.

and even necessary to improve the performance of the (wired) link between the digital subscriber line (DSL) access multiplexors (or central office) and customers' facilities and/or the (wireless) links between some access points and their respective receivers (personal computers, laptops, etc). The mentioned scenarios give a strong motivation for studying the following system composed of two transmitters communicating with their respective receivers and which can use a relay node. The channel model used to analyze this type of network has been called the interference relay channel (IRC) in [3][4] where the authors introduce a channel with two transmitters, two receivers, and one relay, all of them operating in the same frequency band. The main contribution of [3][4] is to derive achievable transmission rate regions for Gaussian IRCs assuming that the relay is implementing the decode-and-forward protocol (DF) and dirty paper coding.

In this paper, we consider multi-band interference relay channels and three different types of protocols at the relay, namely DF, estimate-and-forward (EF), and amplify-and-forward (AF). One of our main objectives is to study the corresponding power allocation (PA) problems at the transmitters. To this end, we proceed in two main steps. First, we provide achievable transmission rates for single-band Gaussian IRCs when DF, EF, and AF are respectively assumed. Second, we use these results to analyze the properties of the transmission rates for the multi-band case. In the multi-band case, we assume that the transmitters are decision makers that can freely choose their own resource allocation policies while selfishly maximizing their transmission rates. This resource allocation problem can be modeled as a static non-cooperative game. The closest works concerning the game-theoretic approach we adopt here seem to be [5][6][7][8] and [9][10][11]. In [5] [6], the authors study the frequency selective and the parallel interference channels and provide sufficient conditions on the channel gains that ensure the existence and uniqueness of the Nash equilibrium (NE) and convergence of iterative water-filling algorithms. These conditions have been further refined in [7]. In [9], a traffic game in parallel relay networks is considered where each source chooses its power allocation policy to minimize a certain cost function. The price of anarchy [12] is analyzed in such a scenario. In [10], a quite similar analysis is conducted for multi-hop networks. In [11], the authors consider a special case of the Gaussian IRC where there are no direct links between the sources and destinations and there are two dedicated relays (one for each source-destination pair) implementing DF. The power allocation game consists in sharing the user's power between the source and relay transmission. The existence, uniqueness of, and convergence to a NE issues are addressed. In the present paper however, we mainly focus on the existence issue of an NE in the games under study, which is already a non-trivial problem. The uniqueness, efficiency, and the design of convergent distributed power allocation algorithms are studied only in a special case and the generalization is left as very useful extension of the present paper.

This paper is structured as follows. Sec. II describes the system model and assumptions in multi-band IRCs. Sec. III provides achievable transmission rates for single-band IRCs. These rates are exploited further in multi-band IRCs (as users' utility functions) analyzed in Sec. IV where the existence issue of NE in the non-cooperative power allocation game is studied. Three relaying protocols are considered: DF, EF, and AF. Sec. IV provides additional results on uniqueness of NE and convergence to NE for the AF protocol. Sec. V illustrates simulations highlighting the importance of optimally locating the relay and the efficiency of the possible NE. We conclude with summarizing remarks and possible extensions in Sec. VI.

## II. System Model

The system under investigation is represented in Fig. 1. It is composed of two source nodes $\mathcal{S}_1$, $\mathcal{S}_2$ (also called transmitters), transmitting their private messages to their respective destination nodes $\mathcal{D}_1$, $\mathcal{D}_2$ (also called receivers). To this end, each source can exploit $Q$ non-overlapping frequency bands (the notation (q) will be used to refer to band $q \in \{1, \dots, Q\}$) which are assumed to be of unit bandwidth. The signals transmitted by $\mathcal{S}_1$ and $\mathcal{S}_2$ in band (q), denoted by $X_1^{(q)}$ and $X_2^{(q)}$, respectively, are assumed to be independent and power constrained:

$$\forall i \in \{1,2\},\ \sum_{q=1}^{Q} \mathbb{E}|X_i^{(q)}|^2 \leq P_i. \tag{1}$$

For $i \in \{1,2\}$, we denote by $\theta_i^{(q)}$ the fraction of power that is used by $\mathcal{S}_i$ for transmitting in band (q) that is, $\mathbb{E}|X_i^{(q)}|^2 = \theta_i^{(q)} P_i$. Additionally, we assume that there exists a multi-band relay $\mathcal{R}$. With these notations, the signals received by $\mathcal{D}_1$, $\mathcal{D}_2$, and $\mathcal{R}$ in band (q) express as:

$$\left\{ \begin{array}{lcl} Y_1^{(q)} & = & h_{11}^{(q)} X_1^{(q)} + h_{21}^{(q)} X_2^{(q)} + h_{r1}^{(q)} X_r^{(q)} + Z_1^{(q)} \\ Y_2^{(q)} & = & h_{12}^{(q)} X_1^{(q)} + h_{22}^{(q)} X_2^{(q)} + h_{r2}^{(q)} X_r^{(q)} + Z_2^{(q)} \\ Y_r^{(q)} & = & h_{1r}^{(q)} X_1^{(q)} + h_{2r}^{(q)} X_2^{(q)} + Z_r^{(q)} \end{array} \right. \tag{2}$$

where $Z_i^{(q)} \sim \mathcal{N}(0, N_i^{(q)})$, $i \in \{1,2,r\}$, represents the Gaussian complex noise on band (q) and, for all $(i,j) \in \{1,2\}^2$, $h_{ij}^{(q)}$ is the channel gain between $\mathcal{S}_i$ and $\mathcal{D}_j$ and $h_{ir}^{(q)}$ is the channel gain between $\mathcal{S}_i$ and $\mathcal{R}$ in band $(q)$. The channel gains are considered to be static. In wireless networks, this would amount, for instance, to considering a realistic situation where only large scale propagation effects can be taken into account by the transmitters to optimize their rates. The proposed approach can be applied to other types of channel models. Concerning channel state information (CSI), we will always assume coherent communications for each transmitter-receiver pair $(\mathcal{S}_i, \mathcal{D}_i)$ whereas, at the transmitters, the information assumptions will be context dependent. The single-user decoding (SUD) will always be assumed at $\mathcal{D}_1$ and $\mathcal{D}_2$. This is a realistic assumption in a framework where devices communicate in an a priori uncoordinated manner. At the relay, the implemented reception scheme will depend on the protocol assumed. The expressions of the signals transmitted by the relay, $X_r^{(q)}, q \in \{1, ..., Q\}$, depend on the relay protocol assumed and will therefore also be explained in the corresponding sections. So far, we have not mentioned any power constraint on the signals $X_r^{(q)}$. Note that the signal model (2) is sufficiently general for addressing two important scenarios. If one imposes an overall power constraint $\sum_{q=1}^{Q} \mathbb{E}|X_r^{(q)}|^2 \leq P_r$, multi-carrier IRCs with a single relay can be studied. On the other hand, if one imposes $\mathbb{E}|X_r^{(q)}|^2 \leq P_r^{(q)}, q \in \{1, ..., Q\}$, multi-band IRCs where a relay is available on each band (the relays are not necessarily co-located) can be studied. In this paper, for simplicity reasons and as a first step towards solving the general problem (where both source and relaying nodes optimize their PA policies) we will assume that the relay implements a fixed power allocation policy between the $Q$ available bands ($\mathbb{E}|X_r^{(q)}|^2 = P_r^{(q)}, q \in \{1, ..., Q\}$).

To conclude this section, we will mention and justify one additional assumption. As in [4][3][13], the relay will be assumed to operate in the full-duplex mode. Mathematically, it is known from [14] that the achievability proofs for the full-duplex case can be almost directly applied the half-duplex case. But this is not our main motivation. Our main motivation is that, in some communication scenarios, the full-duplex assumption is realistic (see e.g., [1] where the transmit and receive radio-frequency parts are not co-located) and even more suited. In the scenario of DSL systems mentioned in Sec. I, the relay is connected to the source and destination through wired links. This allows the implementation of full-duplex repeaters, amplifiers, or digital relays. The same comment can be applied to optical communications.

**Notational conventions**

The capacity function for complex signals is denoted by $C(x) \triangleq \log_2(1+x)$; for all $a \in [0, 1]$, the quantity $\overline{a}$ stands for $\overline{a} = 1 - a$; the notation $-i$ means that $-i = 1$ if $i = 2$ and $-i = 2$ if $i = 1$; for all complex numbers $c \in \mathbb{C}$, $c^*$, $|c|$, $\mathcal{R}e(c)$ and $\mathcal{I}m(c)$ denote the complex conjugate, modulus and the real and imaginary parts respectively.

## III. Achievable transmission rates for single-band IRCs

This section provides preliminary results regarding the achievable rate regions for the IRCs assuming DF, EF, and AF protocols. They are necessary to express transmission rates in the multi-band case. Thus, we do not aim at improving available rate regions for IRCs as in [13] and related works [15][16][17]. In the latter references, the authors consider some special cases of the discrete IRC and derive rate regions based on the DF protocol and different coding-decoding schemes. In what follows, we make some suboptimal choices for the used coding-decoding schemes and relaying protocols which are motivated by a decentralized framework where each destination does not know the codebook used by the other destination. This approach, facilitates the deployment of relays since the receivers do not need to be modified. In particular, this explains why we do not exploit techniques like rate-splitting or successive interference cancellation. As we assume single-band IRCs, we have that $Q = 1$. For the sake of clarity, we omit the superscript $(1)$ from the different quantities used e.g., $X_i^{(1)}$ becomes in this section $X_i$.

### A. Transmission rates for the DF protocol

One of the purposes of this section is to state a corollary from [3]. Indeed, the given result corresponds to the special case of the rate region derived in [3] where each source sends to its respective destination a private message only (and not both public and private messages as in [3]). The reason for providing this region here is threefold: it is necessary for the multi-band case, it is used in the simulation part to establish a comparison between the different relaying protocols under consideration in this paper, and it makes the paper sufficiently self-contained. The principle of the DF protocol is detailed in [14] and we give here only the main idea behind it. Consider a Gaussian relay channel where the source-relay link has a better quality than the source-destination link. From each message intended for the destination, the source builds a coarse and a fine message. With these two messages, the source superposes two codewords. The rates associated with these codewords (or messages) are such that the relay can reliably decode both of them while the destination can only decode the coarse message. After decoding this message, the destination can subtract the corresponding signal and try to decode the fine message. To help the destination to do so, the relay cooperates with the source by sending some information about the fine message. Mathematically, this translates as follows. The signal transmitted by $\mathcal{S}_i$ is structured as $X_i = X_{i0} + \sqrt{\frac{\tau_i}{\nu_i}\frac{P_i}{P_r}}X_{ri}$. The signals $X_{i0}$ and $X_{ri}$ are independent and correspond to the coarse and fine messages respectively; the parameter $\nu_i$ represents the fraction of transmit power the relay allocates to user $i$, hence we have $\nu_1 + \nu_2 \leq 1$; the parameter $\tau_i$ represents the fraction of transmit power $\mathcal{S}_i$ allocates to the cooperation signal (conveying the fine message). Therefore, we have the following result.

*Corollary 3.1 ([3]): When DF is assumed, the region in (3) is achievable; for $i \in \{1, 2\}$, where $j = -i$,*

$$R_i \leq \min\left\{C\left(\frac{|h_{ir}|^2\overline{\tau_i}P_i}{|h_{jr}|^2\overline{\tau_j}P_j+N_r}\right), C\left(\frac{|h_{ii}|^2P_i+|h_{ri}|^2\nu_iP_r+2\mathcal{R}e(h_{ii}h_{ri}^*)\sqrt{\tau_iP_i\nu_iP_r}}{|h_{ji}|^2P_j+|h_{ri}|^2\nu_jP_r+2\mathcal{R}e(h_{ji}h_{ri}^*)\sqrt{\tau_jP_j\nu_jP_r}+N_i}\right)\right\} \quad (3)$$

$(\nu_1, \nu_2) \in [0,1]^2$ *s.t.* $\nu_1 + \nu_2 \leq 1$ *and* $(\tau_1, \tau_2) \in [0,1]^2$, $\tau_1 + \tau_2 \leq 1$.

In a context of decentralized networks, each source $\mathcal{S}_i$ has to optimize the parameter $\tau_i$ in order to maximize its transmission rate $R_i$. In the rate region above, one can observe that this choice is not independent of the choice of the other source. Therefore, each source finds its optimal strategy by optimizing its rate w.r.t. $\tau_i^*(\tau_j)$. In order to do that, each source has to make some assumptions on the value $\tau_j$ used by the other source. This is precisely a non-cooperative game where each player makes some assumptions on the other player's behavior and maximizes its own utility. Interestingly, we see that, even in the single-band case, the DF protocol introduces a power allocation game through the parameter $\tau_i$ representing the cooperation degree between the source $\mathcal{S}_i$ and relay. In this paper, for obvious reasons of space, we will restrict our attention to the case where the cooperation degrees are fixed. In other words, in the multi-band scenario, the transmitter strategy will consist in choosing only the power allocation policy over the available bands. For more details on the game induced by the cooperation degrees the reader is referred to [2].

### B. Transmission rates for the EF protocol

Here, we consider a second main class of relaying protocols, namely the estimate-and-forward protocol. A well-known property of the EF protocol for the relay channel [14] is that it always improves the performance of the receiver w.r.t. the case without relay (in contrast with DF protocols which can degrade the performance of the point-to-point link). The principle of the EF protocol for the standard relay channel is that the relay sends an approximated version of its observation signal to the receiver. More precisely, from an information-theoretic point of view [14], the relay compresses its observation in the Wyner-Ziv manner [18], i.e., knowing that the destination also receives a direct signal from the source that is correlated with the signal to be compressed. The compression rate is precisely tuned by taking into account this correlation degree and the quality of the relay-destination link. In our setup, we have two different receivers. The relay can either create a single quantized version of its observation, common to both receivers, or two quantized versions, one adapted for each destination. We have chosen the second type of quantization which we call the "bi-level compression EF". We note the work by [19] where the authors consider a different channel, namely a

$$R_1 \le C\left(\frac{|h_{11}|^2P_1}{N_1+\frac{|h_{21}|^2P_2\left(N_r+N_{wz}^{(1)}\right)}{|h_{2r}|^2P_2+N_r+N_{wz}^{(1)}}}+\frac{|h_{1r}|^2P_1}{N_r+N_{wz}^{(1)}+\frac{|h_{2r}|^2P_2N_1}{|h_{21}|^2P_2+N_1}}\right), \tag{4}$$

$$R_2 \le C\left(\frac{|h_{22}|^2P_2}{N_2+|h_{r2}|^2\nu_1P_r+\frac{|h_{12}|^2P_1\left(N_r+N_{wz}^{(2)}\right)}{|h_{1r}|^2P_1+N_r+N_{wz}^{(2)}}}+\frac{|h_{2r}|^2P_2}{N_r+N_{wz}^{(2)}+\frac{|h_{1r}|^2P_1(|h_{r2}|^2\nu_1P_r+N_2)}{|h_{12}|^2P_1+|h_{r2}|^2\nu_1P_r+N_2}}\right), \tag{5}$$

separated two-way relay channel, and exploit a similar idea, namely using two quantization levels at the relay. In the scheme used here, each receiver decodes independently its own message, which is less demanding than a joint decoding scheme in terms of information assumptions. As we have already mentioned, the relay implements the Wyner-Ziv compression and superposition coding similarly to a broadcast channel. The difference with the broadcast channel is that each destination also receives the two direct signals from the source nodes. The rate region which can be obtained by using such a coding scheme is given by the following theorem proved in Appendix A.

*Theorem 3.2: For the Gaussian IRC with private messages and bi-level compression EF protocol, any rate pair* $(R_1,R_2)$ *is achievable where*

1) *if* $C\left(\frac{|h_{r1}|^2\nu_2P_r}{|h_{11}|^2P_1+|h_{21}|^2P_2+|h_{r1}|^2\nu_1P_r+N_1}\right)\ge C\left(\frac{|h_{r2}|^2\nu_2P_r}{|h_{22}|^2P_2+|h_{12}|^2P_1+|h_{r2}|^2\nu_1P_r+N_2}\right)$*, we have the rates in (4), (5) subject to the constraints*

$$\begin{cases} N_{wz}^{(1)} \ge \frac{\left(|h_{11}|^2P_1+|h_{21}|^2P_2+N_1\right)A-A_1^2}{|h_{r1}|^2\nu_1P_r} \\ N_{wz}^{(2)} \ge \frac{\left(|h_{22}|^2P_2+|h_{12}|^2P_1+|h_{r2}|^2\nu_1P_r+N_2\right)A-A_2^2}{|h_{r2}|^2\nu_2P_r}, \end{cases}$$

2) *else, if* $C\left(\frac{|h_{r2}|^2\nu_1P_r}{|h_{22}|^2P_2+|h_{12}|^2P_1+|h_{r2}|^2\nu_2P_r+N_2}\right)\ge C\left(\frac{|h_{r1}|^2\nu_1P_r}{|h_{11}|^2P_1+|h_{21}|^2P_1+|h_{r1}|^2\nu_2P_r+N_1}\right)$*, we have the rates in (6), (7) subject to the constraints*

$$\begin{cases} N_{wz}^{(1)} \ge \frac{\left(|h_{11}|^2P_1+|h_{21}|^2P_2+|h_{r1}|^2\nu_2P_r+N_1\right)A-A_1^2}{|h_{r1}|^2\nu_1P_r} \\ N_{wz}^{(2)} \ge \frac{\left(|h_{22}|^2P_2+|h_{12}|^2P_1+N_2\right)A-A_2^2}{|h_{r2}|^2\nu_2P_r}, \end{cases}$$

3) *else the rates are given by (8), (9) subject to the constraints*

$$\begin{cases} N_{wz}^{(1)} \ge \frac{\left(|h_{11}|^2P_1+|h_{21}|^2P_2+|h_{r1}|^2\nu_2P_r+N_1\right)A-A_1^2}{|h_{r1}|^2\nu_1P_r} \\ N_{wz}^{(2)} \ge \frac{\left(|h_{22}|^2P_2+|h_{12}|^2P_1+|h_{r2}|^2\nu_1P_r+N_2\right)A-A_2^2}{|h_{r2}|^2\nu_2P_r}, \end{cases}$$

*with* $N_{wz}^{(i)}$ *representing the quantization noise corresponding to receiver i,* $(\nu_1,\nu_2)\in[0,1]^2, \nu_1+\nu_2\le 1$*, the relay PA,* $A=|h_{1r}|^2P_1+|h_{2r}|^2P_2+N_r$*,* $A_1=2\mathcal{R}e(h_{11}h_{1r}^*)P_1+2\mathcal{R}e(h_{21}h_{2r}^*)P_2$ *and* $A_2=2\mathcal{R}e(h_{12}h_{1r}^*)P_1+2\mathcal{R}e(h_{22}h_{2r}^*)P_2$.

$$R_1 \leq C\left(\frac{|h_{11}|^2 P_1}{N_1 + |h_{r1}|^2 \nu_2 P_r + \frac{|h_{21}|^2 P_2 \left(N_r + N_{wz}^{(1)}\right)}{|h_{2r}|^2 P_2 + N_r + N_{wz}^{(1)}}} + \frac{|h_{1r}|^2 P_1}{N_r + N_{wz}^{(1)} + \frac{|h_{2r}|^2 P_2 (|h_{r1}|^2 \nu_2 P_r + N_1)}{|h_{21}|^2 P_2 + |h_{r1}|^2 \nu_2 P_r + N_1}}\right), \tag{6}$$

$$R_2 \leq C\left(\frac{|h_{22}|^2 P_2}{N_2 + \frac{|h_{12}|^2 P_1 \left(N_r + N_{wz}^{(2)}\right)}{|h_{1r}|^2 P_1 + N_r + N_{wz}^{(2)}}} + \frac{|h_{2r}|^2 P_2}{N_r + N_{wz}^{(2)} + \frac{|h_{1r}|^2 P_1 N_2}{|h_{12}|^2 P_1 + N_2}}\right), \tag{7}$$

$$R_1 \leq C\left(\frac{|h_{11}|^2 P_1}{N_1 + |h_{r1}|^2 \nu_2 P_r + \frac{|h_{21}|^2 P_2 \left(N_r + N_{wz}^{(1)}\right)}{|h_{2r}|^2 P_2 + N_r + N_{wz}^{(1)}}} + \frac{|h_{1r}|^2 P_1}{N_r + N_{wz}^{(1)} + \frac{|h_{2r}|^2 P_2 (|h_{r1}|^2 \nu_2 P_r + N_1)}{|h_{21}|^2 P_2 + |h_{r1}|^2 \nu_2 P_r + N_1}}\right), \tag{8}$$

$$R_2 \leq C\left(\frac{|h_{22}|^2 P_2}{N_2 + |h_{r2}|^2 \nu_1 P_r + \frac{|h_{12}|^2 P_1 \left(N_r + N_{wz}^{(2)}\right)}{|h_{1r}|^2 P_1 + N_r + N_{wz}^{(2)}}} + \frac{|h_{2r}|^2 P_2}{N_r + N_{wz}^{(2)} + \frac{|h_{1r}|^2 P_1 (|h_{r2}|^2 \nu_1 P_r + N_2)}{|h_{12}|^2 P_1 + |h_{r2}|^2 \nu_1 P_r + N_2}}\right), \tag{9}$$

The three scenarios emphasized in this theorem correspond to the following situations: 1) $\mathcal{D}_1$ has the better link (in the sense of the theorem) and can decode both the relay message intended for $\mathcal{D}_2$ and its own message; 2) this scenario is the dual of scenario 1); 3) in this latter scenario, each destination node sees the cooperation signal intended for the other destination node as interference.

### *C. Transmission rates for the AF protocol*

In this section, the relay is assumed to implement an analog amplifier which does not introduce any delay on the relayed signal. The main features of AF-type protocols are well-known by now (e.g., such relays are generally cheap, involve low complexity relay transceivers, and generally induce negligible processing delays in contrast with DF and EF-type relaying protocols). The relay merely sends $X_r = a_r Y_r$ where $a_r$ corresponds to the relay amplification factor/gain. We call the corresponding protocol the zero-delay scalar amplify-and-forward (ZDSAF). The type of assumptions we make here fits well to the setting of DSL or optical communication networks. In wireless networks, the assumed protocol can be seen as an approximation of a scenario with a relay equipped with a power amplifier only. The following theorem provides a region of transmission rates that can be achieved when the transmitters send private messages to their respective receivers, the relay implements the ZDSAF protocol and the receivers implement single-user decoding. The considered framework is attractive in the sense that an AF-based relay can be added to the network without changing the receivers.

*Theorem 3.3 (Transmission rate region for the IRC with ZDSAF): Let* $R_i$, $i \in \{1,2\}$, *be the transmission rate for the source node* $\mathcal{S}_i$. *When ZDSAF is assumed the following region is achievable:*

$$\forall i \in \{1,2\}, R_i^{\mathrm{AF}} \leq C\left(\frac{|a_r \ \ h_{ir}h_{ri}+h_{ii}|^2\rho_i}{|a_r \ \ h_{jr}h_{ri}+h_{ji}|^2\rho_j\frac{N_j}{N_i}+a_r^2|h_{ri}|^2\frac{N_r}{N_i}+1}\right) \tag{10}$$

*where* $\rho_i = \frac{P_i}{N_i}$, $j = -i$, *and* $a_r$ *is the relay amplification gain.*

The proof of this result is standard [20] and will therefore be omitted. Only two points are worth being mentioned. First, the proposed region is achieved by using Gaussian codebooks. Second, the choice of the value of the amplification gain $a_r$ is not always straightforward. In the vast majority of the papers available in the literature, $a_r$ is chosen in order to saturate the power constraint at the relay ($\mathbb{E}|X_r|^2 = P_r$) that is: $a_r = \overline{a}_r \triangleq \sqrt{\frac{P_r}{\mathbb{E}|Y_r|^2}} = \sqrt{\frac{P_r}{|h_{1r}|^2P_1+|h_{2r}|^2P_2+N_r}}$. However, as mentioned in some works [21][22][23][24], this choice can be sub-optimal in the sense of certain performance criteria. The intuitive reason for this is that the AF protocol not only amplifies the useful signal but also the received noise. This effect can be negligible in certain scenarios for the standard relay channel but can be significant for the IRC. Indeed, even if the noise at the relay is negligible, the interference term for user $i$ (i.e., the term $h_{jr}X_j$, $j = -i$) can be influential. In order to assess the importance of choosing amplification factor $a_r$ adequately (i.e., to maximize the transmission rate of a given user or the network sum-rate) we derive its best value. The proposed derivation differs from [21][23] because, here, we consider a different system (an IRC instead of a relay channel with no direct link), a specific relaying function (linear relaying functions instead of arbitrary functions) and a different performance metric (individual transmission rate and sum-rate instead of raw bit error rate [21] and mutual information [23]). Our problem is also different from [24] since we do not consider the optimal clipping threshold in the sense of the end-to-end distortion for frequency division relay channels. At last, the main difference with [22] is that, for the relay channel, the authors discuss the choice of the optimal amplification gain in terms of transmission rate for a vector AF protocol having a delay of at least one symbol duration; here we focus on a scalar AF protocol with no delay and a different system namely the IRC. In this setup, we have found an analytical expression for the best $a_r$ in the sense of $R_i(a_r)$ for a given user $i \in \{1,2\}$. We have also noticed that the $a_r$ maximizing the network sum-rate has to be computed numerically in general. The corresponding analytical result is stated in the following theorem.

*Theorem 3.4:* [Optimal amplification gain for the ZDSAF in the IRC] *The transmission rate of user* $i$*,* $R_i(a_r)$*, as a function of* $a_r \in [0, \overline{a}_r]$ *can have several critical points which are the real solutions, denoted by* $c_{r,i}^{(1)}$ *and* $c_{r,i}^{(2)}$*, to the following second degree equation:*

$$\begin{array}{l} a_r^2 \left[|m_i|^2 \mathrm{Re}(p_i q_i^*) - (|p_i|^2 + s_i)\mathrm{Re}(m_i n_i^*)\right] + a_r \left[|m_i|^2(|q_i|^2+1) - |n_i|^2(|p_i|^2+s_i)\right] \\ +(|q_i|^2+1)\mathrm{Re}(m_i n_i^*) - |n_i|^2 \mathrm{Re}(p_i q_i^*) = 0 \end{array} \quad (11)$$

*where* $m_i = h_{ir} h_{ri} \sqrt{\rho_i}$, $n_i = h_{ii}\sqrt{\rho_i}$, $p_i = h_{jr} h_{ri}\sqrt{\rho_j}$, $q_i = h_{ji}\sqrt{\rho_j}$, $s_i = |h_{ri}|^2$, $i \in \{1,2\}$ *and* $j = -i$*. Thus, depending on the channel parameters, the optimal amplification gain* $a_r^* = \arg\max_{a_r \in [0,\overline{a}_r]} R_i(a_r)$ *takes one value in the set* $a_r^* \in \{0, \overline{a}_r, c_{r,i}^{(1)}, c_{r,i}^{(2)}\}$*. If, additionally, the channel gains are reals then the two critical points write as:* $c_{r,i}^{(1)} = -\frac{n_i}{m_i}$ *and* $c_{r,i}^{(2)} = -\frac{m_i q_i^2 + m_i - p_i q_i n_i}{m_i q_i p_i - p_i^2 n_i - n_i s_i}$.

The proof of this result is provided in Appendix B. Of course, in practice, if the receive signal-to-noise plus interference ratio (viewed from a given user) at the relay is low, choosing the amplification factor $a_r$ adequately does not solve the problem. It is well known that in real systems, a more efficient way to combat noise is to implement error correcting codes. This is one of the reasons why DF is also an important relaying protocol, especially for digital relay transceivers for which AF cannot be implemented in its standard form (see e.g., [24] for more details).

### *D. Time-Sharing*

In terms of achievable Shannon rates, distributed channels differ from their centralized counterpart. Indeed, rate regions are not necessarily convex since the time-sharing argument can be invalid (if no synchronization is possible). Similarly, depending on the channel gains, the achievable rate for a given transmitter can be non-concave with respect to its power allocation policy. This happens if the transmitters cannot be coordinated (distributed channels).

Assuming that the users can be coordinated, we discuss here a standard time-sharing procedure similarly to the approach in [25] for the frequency-division relay channel. More specifically, we assume that user 1 decides to transmit only during a fraction $\alpha_1$ of the time using the power $\frac{P_1}{\alpha_1}$ and user 2 transmits only with a fraction $\alpha_2$ percent of the time using the power $\frac{P_2}{\alpha_2}$.

The achievable rate-region with coordinated time-sharing, irrespective of the relay protocol, is:

$$\forall i \in \{1,2\}, R_i^{\mathrm{TS}} \leq \alpha_i \overline{\beta_j} R_i \left(\frac{P_i}{\alpha_i}, 0\right) + \alpha_i \beta_j R_i \left(\frac{P_i}{\alpha_i}, \frac{P_j}{\alpha_j}\right), \tag{12}$$

where $j = -i$, $(\alpha_i, \alpha_j)^2 \in [0,1]^2$, $(\beta_i, \beta_j)^2 \in [0,1]^2$ such that $\beta_1 \alpha_2 = \beta_2 \alpha_1$. The rate $R_i\left(\frac{P_i}{\alpha_i}, 0\right)$ represents the achievable rate of user $i$ (depends on the relay protocol and was provided in the previous subsections) when user $j$ doesn't transmit and user $i$ transmits with power $\frac{P_i}{\alpha_i}$, $R_i\left(\frac{P_i}{\alpha_i}, \frac{P_j}{\alpha_j}\right)$ is the achievable rate when user $i$ transmits with power $\frac{P_i}{\alpha_i}$ and user $j$ transmits with power $\frac{P_j}{\alpha_j}$. In order to achieve this rate region, the users have to be coordinated. This means that they have to know at each instant if the other user is transmitting or not. User $i$ also has to know the parameters $\alpha_i$ and $\alpha_j$. The parameter $\beta_j \in [0,1]$ represents the fraction of time when user $j$ interferes with user $i$. Considering the time when both users transmit with non-zero power, we obtain the condition: $\beta_1 \alpha_2 = \beta_2 \alpha_1$.

## IV. Power allocation games in multi-band IRCs and Nash equilibrium analysis

In the previous section, we have considered the system model presented in Sec. II for $Q = 1$. Here, we consider multi-band IRCs for which $Q \geq 2$. As communications interfere on each band, choosing the power allocation policy at a given transmitter is not a simple optimization problem. Indeed, this choice depends on what the other transmitter does. Each transmitter is assumed to optimize its transmission rate in a selfish manner by allocating its transmit power $P_i$ between $Q$ sub-channels and knowing that the other transmitters want to do the same. This interaction can be modeled as a strategic form non-cooperative game, $\mathcal{G} = (\mathcal{K}, (\mathcal{A}_i)_{i \in \mathcal{K}}, (u_i)_{i \in \mathcal{K}})$, where: (i) the *players* of the game are the two information sources or transmitters and $\mathcal{K} = \{1,2\}$ is used to refer to the set of players; (ii) the strategy of transmitter $i$ consists in choosing $\underline{\theta}_i = (\theta_i^{(1)}, \ldots, \theta_i^{(Q)})$ in its *strategy set* $\mathcal{A}_i = \left\{ \underline{\theta}_i \in [0,1]^Q \,\middle|\, \sum_{q=1}^{Q} \theta_i^{(q)} \leq 1 \right\}$ where $\theta_i^{(q)}$ represents the fraction of power used in band $(q)$; (iii) $u_i(\cdot)$ is the *utility* function of user $i \in \{1,2\}$ or its achievable rate depending on the relaying protocol. From now on, we will call *state* of the network the (concatenated) vector of power fractions that the transmitters allocate to the IRCs i.e., $\underline{\theta} = (\underline{\theta}_1, \underline{\theta}_2)$. An important issue is to determine whether there exist some outcomes to this conflicting situation. A natural solution concept in non-cooperative games is the Nash equilibrium [26]. In distributed networks, the existence of a stable operating state of the system is a desirable feature. In this respect, the NE is a stable state from which the users do not have any incentive to unilaterally deviate (otherwise they would lose in terms of utility). The mathematical definition is the following.

$$\begin{cases} R_{1,1}^{(q),\mathrm{DF}} &= C\left(\frac{\left|h_{1r}^{(q)}\right|^2\overline{\tau_1^{(q)}}\theta_1^{(q)}P_1}{\left|h_{2r}^{(q)}\right|^2\overline{\tau_2^{(q)}}\theta_2^{(q)}P_2+N_r^{(q)}}\right) \\ R_{2,1}^{(q),\mathrm{DF}} &= C\left(\frac{\left|h_{2r}^{(q)}\right|^2\overline{\tau_2^{(q)}}\theta_2^{(q)}P_2}{\left|h_{1r}^{(q)}\right|^2\overline{\tau_1^{(q)}}\theta_1^{(q)}P_1+N_r^{(q)}}\right) \\ R_{1,2}^{(q),\mathrm{DF}} &= C\left(\frac{\left|h_{11}^{(q)}\right|^2\theta_1^{(q)}P_1+\left|h_{r1}^{(q)}\right|^2\nu^{(q)}P_r^{(q)}+2\mathrm{Re}\left(h_{11}^{(q)}h_{r1}^{(q),*}\right)\sqrt{\tau_1^{(q)}\theta_1^{(q)}P_1\nu^{(q)}P_r^{(q)}}}{\left|h_{21}^{(q)}\right|^2\theta_2^{(q)}P_2+\left|h_{r1}^{(q)}\right|^2\overline{\nu^{(q)}}P_r^{(q)}+2\mathrm{Re}\left(h_{21}^{(q)}h_{r1}^{(q),*}\right)\sqrt{\tau_2^{(q)}\theta_2^{(q)}P_2\overline{\nu^{(q)}}P_r^{(q)}}+N_1^{(q)}}\right) \\ R_{2,2}^{(q),\mathrm{DF}} &= C\left(\frac{\left|h_{22}^{(q)}\right|^2\theta_2^{(q)}P_2+\left|h_{r2}^{(q)}\right|^2\overline{\nu^{(q)}}P_r^{(q)}+2\mathrm{Re}\left(h_{22}^{(q)}h_{r2}^{(q),*}\right)\sqrt{\tau_2^{(q)}\theta_2^{(q)}P_2\overline{\nu^{(q)}}P_r^{(q)}}}{\left|h_{12}^{(q)}\right|^2\theta_1^{(q)}P_1+\left|h_{r2}^{(q)}\right|^2\nu^{(q)}P_r^{(q)}+2\mathrm{Re}\left(h_{12}^{(q)}h_{r2}^{(q),*}\right)\sqrt{\tau_1^{(q)}\theta_1^{(q)}P_1\nu^{(q)}P_r^{(q)}}+N_2^{(q)}}\right), \end{cases} \tag{14}$$

*Definition 4.1:* [Nash equilibrium] *The state* $(\underline{\theta}_i^*, \underline{\theta}_{-i}^*)$ *is a pure NE of the strategic form game* $\mathcal{G}$ *if* $\forall i \in \mathcal{K}, \forall \underline{\theta}_i' \in \mathcal{A}_i,\ u_i(\underline{\theta}_i^*, \underline{\theta}_{-i}^*) \geq u_i(\underline{\theta}_i', \underline{\theta}_{-i}^*)$.

In this section, we mainly focus on the problem of existence of such a solution, which is the first step towards equilibria characterization in IRCs. The problems of equilibrium uniqueness, selection, convergence, and efficiency are therefore left as natural extensions of the work reported here.

### A. Equilibrium existence analysis for the DF protocol

As explained in Sec. III-A the signals transmitted by $\mathcal{S}_1$ and $\mathcal{S}_2$ in band $(q)$ have the following form: $X_i^{(q)} = X_{i,0}^{(q)} + \sqrt{\frac{\tau_i^{(q)}}{\nu_i^{(q)}}\frac{\theta_i^{(q)}P_i}{P_r^{(q)}}}X_{r,i}^{(q)}$ where the signals $X_{i,0}^{(q)}$ and $X_{r,i}^{(q)}$ are Gaussian and independent. At the relay $\mathcal{R}$, the transmitted signal writes as: $X_r^{(q)} = X_{r,1}^{(q)} + X_{r,2}^{(q)}$. For a given allocation policy $\underline{\theta}_i = \left(\theta_i^{(1)}, ..., \theta_i^{(Q)}\right)$, the source-destination pair $(\mathcal{S}_i, \mathcal{D}_i)$ achieves the transmission rate $\sum_{q=1}^{Q} R_i^{(q),\mathrm{DF}}$ where

$$\begin{cases} R_1^{(q),\mathrm{DF}} &= \min\left\{R_{1,1}^{(q),\mathrm{DF}}, R_{1,2}^{(q),\mathrm{DF}}\right\} \\ R_2^{(q),\mathrm{DF}} &= \min\left\{R_{2,1}^{(q),\mathrm{DF}}, R_{2,2}^{(q),\mathrm{DF}}\right\} \end{cases}, \tag{13}$$

and $R_{1,1}^{(q),\mathrm{DF}}$, $R_{1,2}^{(q),\mathrm{DF}}$,$R_{2,1}^{(q),\mathrm{DF}}$,$R_{2,2}^{(q),\mathrm{DF}}$ are given in (14) and $(\nu^{(q)}, \tau_1^{(q)}, \tau_2^{(q)})$ is a given triple of parameters in $[0,1]^3$, $\tau_1^{(q)} + \tau_2^{(q)} \leq 1$.

The achievable transmission rate of user $i$ is given by:

$$u_i^{\mathrm{DF}}(\underline{\theta}_i, \underline{\theta}_{-i}) = \sum_{q=1}^{Q} R_i^{(q),\mathrm{DF}}(\theta_i^{(q)}, \theta_{-i}^{(q)}). \tag{15}$$

We suppose that the game is played once (one-shot or static game), the users are rational (each selfish player does what is best for itself), rationality is common knowledge, and the game is with complete information that is, every player knows the triplet $\mathcal{G}^{\mathrm{DF}} = (\mathcal{K}, (\mathcal{A}_i)_{i\in\mathcal{K}}, (u_i^{\mathrm{DF}})_{i\in\mathcal{K}})$. Although this setup might seem to be demanding in terms of CSI at the source nodes, it turns out that the equilibria predicted in such a framework can be effectively observed in more realistic frameworks where one player observes the strategy played by the other player and reacts accordingly by maximizing his utility, the other player observes this and updates its strategy and so on. We will come back to this later on. The existence theorem for the DF protocol is given hereunder.

*Theorem 4.2:* [Existence of an NE for the DF protocol]
*If the channel gains satisfy the condition* $\mathcal{R}e(h_{ii}^{(q)} h_{ri}^{(q)*}) \geq 0$*, for all* $i \in \{1, 2\}$ *and* $q \in \{1, \dots, Q\}$ *the game defined by* $\mathcal{G}^{\mathrm{DF}} = (\mathcal{K}, (\mathcal{A}_i)_{i\in\mathcal{K}}, (u_i^{\mathrm{DF}}(\underline{\theta}_i, \underline{\theta}_{-i}))_{i\in\mathcal{K}})$ *with* $\mathcal{K} = \{1, 2\}$ *and* $\mathcal{A}_i = \left\{ \underline{\theta}_i \in [0, 1]^Q \,\middle|\, \sum_{q=1}^{Q} \theta_i^{(q)} \leq 1 \right\}$*, has always at least one pure NE.*

*Proof:* The proof is based on Theorem 1 of [27]. The latter theorem states that in a game with a finite number of players, if for every player 1) the strategy set is convex and compact, 2) its utility is continuous in the vector of strategies and 3) concave in its own strategy, then the existence of at least one pure NE is guaranteed. In our setup checking that conditions 1) and 2) are met is straightforward. The condition $\mathcal{R}e(h_{ii}^{(q)} h_{ri}^{(q)*}) \geq 0$ is a sufficient condition that ensures the concavity of $R_{i,2}^{\mathrm{DF}}$ w.r.t. $\theta_i^{(q)}$. The intuition behind this condition is that the superposition of the two signals carrying useful information for user $i$ (i.e., signal from $\mathcal{S}_i$ and $\mathcal{R}$) has to be constructive w.r.t. the amplitude of the resulting signal. As the sum of concave functions is a concave function, the *min* of two concave functions is a concave function (see [28] for more details on operations preserving concavity), and $R_{i,j}^{(q)}$ is a concave function of $\underline{\theta}_i$, it follows that 3) is also met, which concludes the proof. ∎

Theorem indicates, in particular, that for the pathloss model where the channel gains are positive real scalars (i.e., $h_{ij} > 0$, $(i, j) \in \{1, 2, r\}^2$) there always exists an equilibrium. As a consequence, if some relays are added in the network, the transmitters will adapt their PA policies accordingly and, whatever the locations of the relays, an equilibrium will be observed. This is a nice property for the system under investigation. As the PA game with DF is concave it is tempting to try to verify whether the sufficient condition for uniqueness of [27] is met here. It turns out that the diagonally strict concavity condition of [27] is not trivial to be checked. Additionally, it is possible that the game has several equilibria as it is proven to be the case for the AF protocol.

$$\left\{\begin{array}{lll} R_1^{(q),\mathrm{EF}} & = & C\left(\frac{\left(\left|h_{2r}^{(q)}\right|^2\theta_2^{(q)}P_2+N_r^{(q)}+N_{wz,1}^{(q)}\right)\left|h_{11}^{(q)}\right|^2\theta_1^{(q)}P_1+\left(\left|h_{21}^{(q)}\right|^2\theta_2^{(q)}P_2+\left|h_{r1}^{(q)}\right|^2\overline{\nu^{(q)}}P_r^{(q)}+N_1^{(q)}\right)\left|h_{1r}^{(q)}\right|^2\theta_1^{(q)}P_1}{\left(N_r^{(q)}+N_{wz,1}^{(q)}\right)\left(|h_{21}^{(q)}|^2\theta_2^{(q)}P_2+|h_{r1}^{(q)}|^2\overline{\nu^{(q)}}P_r^{(q)}+N_1^{(q)}\right)+|h_{2r}^{(q)}|^2\theta_2^{(q)}P_2\left(|h_{r1}^{(q)}|^2\overline{\nu^{(q)}}P_r^{(q)}+N_1^{(q)}\right)}\right) \\ R_2^{(q),\mathrm{EF}} & = & C\left(\frac{\left(|h_{1r}|^2\theta_1^{(q)}P_1+N_r^{(q)}+N_{wz,2}^{(q)}\right)\left|h_{22}^{(q)}\right|^2\theta_2^{(q)}P_2+\left(\left|h_{12}^{(q)}\right|^2\theta_1^{(q)}P_1+\left|h_{r2}^{(q)}\right|^2\nu^{(q)}P_r^{(q)}+N_2^{(q)}\right)\left|h_{2r}^{(q)}\right|^2\theta_2^{(q)}P_2}{\left(N_r^{(q)}+N_{wz,2}^{(q)}\right)\left(\left|h_{12}^{(q)}\right|^2\theta_1^{(q)}P_1+\left|h_{r2}^{(q)}\right|^2\nu^{(q)}P_r^{(q)}+N_2^{(q)}\right)+\left|h_{1r}^{(q)}\right|^2\theta_1^{(q)}P_1\left(\left|h_{r2}^{(q)}\right|^2\nu^{(q)}P_r^{(q)}+N_2^{(q)}\right)}\right) \end{array}\right., \tag{17}$$

$$\left\{\begin{array}{lll} N_{wz,1}^{(q)} & = & \frac{\left(\left|h_{11}^{(q)}\right|^2\theta_1^{(q)}P_1+\left|h_{21}^{(q)}\right|^2\theta_2^{(q)}P_2+\left|h_{r1}^{(q)}\right|^2\overline{\nu^{(q)}}P_r^{(q)}+N_1^{(q)}\right)A^{(q)}-\left|A_1^{(q)}\right|^2}{\left|h_{r1}^{(q)}\right|^2\nu^{(q)}P_r^{(q)}} \\ N_{wz,2}^{(q)} & = & \frac{\left(\left|h_{22}^{(q)}\right|^2\theta_2^{(q)}P_2+\left|h_{12}^{(q)}\right|^2\theta_1^{(q)}P_1+\left|h_{r2}^{(q)}\right|^2\nu^{(q)}P_r^{(q)}+N_2^{(q)}\right)A^{(q)}-\left|A_2^{(q)}\right|^2}{\left|h_{r2}^{(q)}\right|^2\overline{\nu^{(q)}}P_r^{(q)}} \end{array}\right. \tag{18}$$

### *B. Equilibrium existence analysis for the EF protocol*

In this section, we make the same assumptions as in Sec. IV-A concerning the reception schemes and PA policies at the relays: we assume that each node $\mathcal{R}$, $\mathcal{D}_1$ and $\mathcal{D}_2$ implements single-user decoding and the PA policy at each relay i.e., $\underline{\nu} = \left(\nu^{(1)}, ..., \nu^{(Q)}\right)$ is fixed. Each relay now implements the EF protocol. Under this assumption, the utility for user $i \in \{1, 2\}$ can be expressed as follows

$$u_i^{\mathrm{EF}}(\underline{\theta}_i, \underline{\theta}_{-i}) = \sum_{q=1}^{Q} R_i^{(q),\mathrm{EF}} \tag{16}$$

where $R_i^{(q),\mathrm{EF}}$ are given in (17) $\nu^{(q)} \in [0, 1]$, $A^{(q)} = |h_{1r}^{(q)}|^2\theta_1^{(q)}P_1+|h_{2r}^{(q)}|^2\theta_2^{(q)}P_2+N_r^{(q)}$, $A_1^{(q)} = h_{11}^{(q)}h_{1r}^{(q),*}\theta_1^{(q)}P_1+h_{21}^{(q)}h_{2r}^{(q),*}\theta_2^{(q)}P_2$ and $A_2^{(q)} = h_{12}^{(q)}h_{1r}^{(q),*}\theta_1^{(q)}P_1 + h_{22}^{(q)}h_{2r}^{(q),*}\theta_2^{(q)}P_2$. What is interesting with this EF protocol is that, here again, one can prove that the utility is concave for every user. This is the purpose of the following theorem.

*Theorem 4.3:* [Existence of an NE for the bi-level compression EF protocol] *The game defined by* $\mathcal{G}^{\mathrm{EF}} = (\mathcal{K}, (\mathcal{A}_i)_{i\in\mathcal{K}}, (u_i^{\mathrm{EF}}(\underline{\theta}_i, \underline{\theta}_{-i}))_{i\in\mathcal{K}})$ *with* $\mathcal{K} = \{1, 2\}$ *and* $\mathcal{A}_i = \left\{\underline{\theta}_i \in [0, 1]^Q \,\middle|\, \sum_{q=1}^{Q} \theta_i^{(q)} \leq 1\right\}$*, has always at least one pure NE.*

The proof is similar to the proof of Theorem IV-A. To be able to apply Theorem $1$ of Rosen [27], we have to prove that the utility $u_i^{\mathrm{EF}}$ is concave w.r.t. $\underline{\theta}_i$. The problem is less simple than for DF because the compression noise $N_{wz,i}^{(q)}$ which appears in the denominator of the capacity function in Eq. (17) depends on the strategy $\underline{\theta}_i$ of transmitter $i$. It turns out that it is still possible to prove the desired result as shown in Appendix C.

$$R_i^{(q),\mathrm{AF}} = C\left(\frac{\left| h_{ir}^{(q)} h_{ri}^{(q)}\right|^2 \theta_i^{(q)} \rho_i \rho_r \frac{N_r}{N_i}}{\left|h_{ri}^{(q)}\right|^2 \theta_i^{(q)} \rho_i + \left(\left|h_{rj}^{(q)}\right|^2 \theta_j^{(q)} \rho_j \frac{N_j}{N_i} + \frac{N_r}{N_i}\right)\left(\left|h_{ri}^{(q)}\right|^2 \rho_r \frac{N_r}{N_i} + 1\right)}\right) \quad (21)$$

### C. Equilibrium analysis for the AF protocol

Here, we assume that the relay implements the ZDSAF protocol, which has already been described in Sec. III-C. One of the nice features of the (analog) ZDSAF protocol is that relays are very easy to be deployed since they can be used without any change on the existing (non-cooperative) communication system. The amplification factor/gain for the relay on band $(q)$ will be denoted by $a_r^{(q)}$. Here, we consider the most common choice for the amplification factor that it, the one that exploits all the transmit power available on each band. The achievable transmission rate is given by

$$u_i^{\mathrm{AF}}(\underline{\theta}_i, \underline{\theta}_{-i}) = \sum_{q=1}^{Q} R_i^{(q),\mathrm{AF}}(\theta_i^{(q)}, \theta_{-i}^{(q)}) \quad (19)$$

where $R_i^{(q),\mathrm{AF}}$ is the rate user $i$ obtains by using band (q) when the ZDSAF protocol is used by the relay $\mathcal{R}$. After Sec. III-C the latter quantity is:

$$\forall i \in \{1,2\}, R_i^{(q),\mathrm{AF}} = C\left(\frac{\left|a_r^{(q)} \ h_{ir}^{(q)} h_{ri}^{(q)} + h_{ii}^{(q)}\right|^2 \theta_i^{(q)} \rho_i}{\left|a_r^{(q)} \ h_{jr} h_{ri} + h_{ji}\right|^2 \rho_j \theta_j^{(q)} \frac{N_j^{(q)}}{N_i^{(q)}} + \left(a_r^{(q)}\right)^2 \left|h_{ri}^{(q)}\right|^2 \frac{N_r^{(q)}}{N_i^{(q)}} + 1}\right), \quad (20)$$

where $a_r^{(q)} = \tilde{a}_r^{(q)}(\theta_1^{(q)}, \theta_2^{(q)}) \triangleq \sqrt{\frac{P_r}{\left|h_{1r}^{(q)}\right|^2 P_1 + |h_{2r}|^2 P_2 + N_r}}$ and $\rho_i^{(q)} = \frac{P_i}{N_i^{(q)}}$. Without loss of generality and for the sake of clarity we will assume in Sec. IV-C that $\forall (i,q) \in \{1,2,r\} \times \{1,\dots,Q\}, N_i^{(q)} = N$, $P_r^{(q)} = P_r$ and we introduce the quantities $\rho_i = \frac{P_i}{N}$. In this setup the following existence theorem can be proven.

*Theorem 4.4:* [Existence of an NE for ZDSAF] *If any of the following conditions are met: i)* $\left|a_r^{(q)} h_{ir}^{(q)} h_{ri}^{(q)}\right| \gg \left|h_{ii}^{(q)}\right|$ *and* $\left|a_r^{(q)} h_{jr}^{(q)} h_{ri}^{(q)}\right| \gg \left|h_{ji}^{(q)}\right|$ *(negligible direct links), ii)* $\left|h_{ii}^{(q)}\right| \gg \left|a_r^{(q)} h_{ir}^{(q)} h_{ri}^{(q)}\right|$ *and* $\left|h_{ji}^{(q)}\right| \gg \left|a_r^{(q)} h_{jr}^{(q)}\right| \min\left\{1, \left|h_{ri}^{(q)}\right|\right\}$ *(negligible relay links), iii)* $a_r^{(q)} = A_r^{(q)} \in [0, \tilde{a}_r^{(q)}(1,1)]$ *(constant amplification gain), there exists at least one pure NE in the PA game* $\mathcal{G}^{\mathrm{AF}}$.

The proof is similar to the proof of Theorem IV-A. The sufficient conditions ensure the concavity of the function $R_i^{(q),\mathrm{AF}}$ w.r.t. $\theta_i^{(q)}$. For the first case i) where the direct links between the sources and destinations are negligible (e.g., in the wired DSL setting these links are missing and the transmission is only possible using the relay nodes), the achievable rates are given $\forall i \in \{1,2\}$, in (21) and it can be proven that $R_i^{(q),\mathrm{AF}}$ is concave w.r.t. $\theta_i^{(q)}$. The other two cases are easier to prove since the amplification gain is either constant or not taken into account and the rate $R_i^{(q),\mathrm{AF}}$ is a composition of a logarithmic function and a linear function of $\theta_i^{(q)}$ and thus concave.

The determination of NE and the convergence issue to one of the NE are far from being trivial in this case. For example, potential games [29] and supermodular games [30] are known to have attractive convergence properties. It can be checked that, the PA game under investigation is neither a potential nor a supermodular game in general. To be more precise, it is a potential game for a set of channel gains which corresponds to a scenario with probability zero (e.g., the parallel multiple access channel). The authors of [31] studied supermodular games for the interference channel with $K = 2$, $Q = 3$, assuming that only one band is shared by the users (IC) while the other bands are private (one interference-free band for each user). Therefore, each user allocates its power between two bands. Their strategies are designed such that the game has strategic complementarities. However, as stated in [31], this design trick does not work for more than two players or if the users can access more than two frequency bands. In conclusion, general convergence results seem to require more advanced tools and further investigations.

**Special case study**

As we have just mentioned, the uniqueness/convergence/efficiency analysis of NE for the DF and EF protocols requires a separate work to be treated properly. However, it is possible to obtain relatively easy some interesting results in a special case of the AF protocol. The reason for analyzing this special case is threefold: a) it corresponds to a possible scenario in wired communication networks; b) it allows us to introduce some game-theoretic concepts that can be used to treat more general cases and possibly the DF and EF protocols; c) it allows us to have more insights on the problem with a more general choice for $a_r^{(q)}$. The special case under investigation is as follows: $Q = 2$ and $\forall q \in \{1, 2\}$, $a_r^{(q)} = A_r^{(q)} \in [0, \tilde{a}_r(1, 1)]$ are constant w.r.t. $\underline{\theta}$. We observe that the strategy set of user $i$ are scalar spaces $\theta_i \in [0, 1]$ because we can consider $\theta_i^{(1)} = \theta_i$ and $\theta_i^{(2)} = \overline{\theta_i}$. For the sake of clarity, we denote by $h_{ij} = h_{ij}^{(1)}$ and $g_{ij} = h_{ij}^{(2)}$. Note that the case $a_r^{(q)} = A_r^{(q)}$ can also be seen as an interference channel for which there is an additional degree of freedom on each band. The choice $Q = 2$ is totally relevant in scenarios where the spectrum is divided in two bands, one shared band where communications interfere and one protected band where they do not (see e.g., [32]). The choice $a_r^{(q)} = const.$ has the advantage of being mathematically simple and allows us to initialize the uniqueness/convergence analysis. Moreover, it corresponds to a suitable model for an analog repeater in the linear regime in wired networks or, more generally, to a power amplifier for which neither automatic gain control is available nor received power estimation mechanism. By making these two assumptions, it is possible to determine exactly the number of Nash equilibria through the notion of best response (BR) functions. The BR of player $i$ to player $j$ is defined by $\mathrm{BR}_i(\theta_j) = \arg\max_{\theta_i} u_i(\theta_i, \theta_j)$. In general, it is a correspondence but in our case it is just a function. The equilibrium points are the intersection points of the BRs of the two players. In this case, using the Lagrangian functions to impose the power constraint, it can be checked that:

$$\mathrm{BR}_i(\theta_j) = \left| \begin{array}{ll} F_i(\theta_j) & \text{if } 0 < F_i(\theta_j) < 1 \\ 1 & \text{if } F_i(\theta_j) \geq 1 \\ 0 & \text{otherwise} \end{array} \right. \tag{22}$$

where $j = -i$, $F_i(\theta_j) \triangleq -\frac{c_{ij}}{c_{ii}}\theta_j + \frac{d_i}{c_{ii}}$ is an affine function of $\theta_j$ for $(i,j) \in \{(1,2),(2,1)\}$, $c_{ii} = 2|A_r^{(1)} h_{ri} h_{ir} + h_{ii}|^2 |A_r^{(2)} g_{ri} g_{ir} + g_{ii}|^2 \rho_i$; $c_{ij} = |A_r^{(1)} h_{ri} h_{ir} + h_{ii}|^2 |A_r^{(2)} g_{ri} g_{jr} + g_{ji}|^2 \rho_j + |A_r^{(1)} h_{ri} h_{jr} + h_{ji}|^2 |A_r^{(2)} g_{ri} g_{ir} + g_{ii}|^2 \rho_j$; $d_i = |A_r^{(1)} h_{ri} h_{ir} + h_{ii}|^2 [|A_r^{(2)} g_{ri} g_{ir} + g_{ii}|^2 \rho_i + |A_r^{(2)} g_{ri} g_{jr} + g_{ji}|^2 \rho_j + A_r^{(2)} |g_{ri}|^2 + 1] - |A_r^{(2)} g_{ri} g_{ir} + g_{ii}|^2 (A_r^{(1)} |h_{ri}|^2 + 1)$. By studying the intersection points between $\mathrm{BR}_1$ and $\mathrm{BR}_2$, one can prove the following theorem (the proof is provided in Appendix D).

*Theorem 4.5 (Number of Nash equilibria for ZDSAF): For the game $\mathcal{G}^{\mathrm{AF}}$ with fixed amplification gains at the relays, (i.e., $\frac{\partial a_r}{\partial \theta_i^{(q)}} = 0$), there can be a unique NE, two NE, three NE or an infinite number of NE, depending on the channel parameters (i.e., $h_{ij}$, $g_{ij}$, $\rho_i$, $A_r^{(q)}$, $(i,j) \in \{1,2,r\}^2$, $q \in \{1,2\}$.*

Notice that, if $A_r = 0$, we obtain the complete characterization of the NE set for the two-users two-channels parallel interference channel. In the proof in Appendix D, we give explicit expressions of the possible NE in function of the system parameters (i.e., the amplification gain $A_r$ and the channel gains). If the channel gains are the realizations of continuous random variables, it is easy to prove that the probability of observing the necessary conditions on the channel gains for having two NEs or an infinite number of NEs is zero. Said otherwise, considering the pathloss model and arbitrary nodes positioning, there will be, with probability one, either one or three NE, depending on the channel gains. When the channel gains are such that the NE is unique, the unique NE can be shown to be:

$$\underline{\theta}^{\mathrm{NE}} = \underline{\theta}^* = \left( \frac{c_{22} d_1 - c_{12} d_2}{c_{11} c_{22} - c_{12} c_{21}}, \frac{c_{11} d_2 - c_{21} d_1}{c_{11} c_{22} - c_{12} c_{21}} \right). \tag{23}$$

When there are three NE, it seems a priori impossible to predict the NE that will be effectively observed in the one-shot game. In fact, in practice, in a context of adaptive/cognitive transmitters (note that what can be adapted is also the PA policy chosen by the designer/owner of the transmitter), it is possible to predict the equilibrium of the network. First, in general, there is no reason why the sources should start transmitting at the same time. Thus, one transmitter, say $i$, will be alone and using a certain PA policy. The transmitter coming after, namely $\mathcal{S}_{-i}$, will sense/measure/probe its environment and play its BR to what it observes. As a consequence, user $i$ will move to a new policy, maximizing its utility to what transmitter $-i$ has played and so on. The key question is: does this procedure converge? This procedure is guaranteed to converge to one of the NE and a detailed discussion about the asymptotic stability of the NE can be found in Appendix D. The arguments for proving this have been used for the first time in [33] where the "Cournot duopoly" was introduced. In [33], the BRs of each player is purely affine, which leads in this case to a unique equilibrium. The corresponding iterative procedure is called the Cournot tâtonnement process in [34]. In the case with three NE, the effectively observed NE can be predicted by knowing the initial network state that is, the PA policy played by the first transmitting player (see Sec. V). To implement such an iterative procedure, it can be checked [1] that the transmitters need to know less network parameters than in the original game where the amplification factor saturates the constraint. In fact, the needed parameters can be acquired by realistic sensing/probing techniques or feedback mechanisms based on standard estimation procedures. As a comment, note that in the (modern) literature of decentralized or distributed communications networks where the optimal PA policy of a transmitter is to water-fill, the mentioned iterative procedure is called iterative water-filling.

### *D. Equilibrium analysis for the Time-Sharing scheme*

In the previous subsections, we have given sufficient conditions that ensure the existence of the Nash equilibrium. Our approach is based on the concave games studied in [27] and consists in finding the sufficient conditions that ensure the concavity of the transmission achievable rates. We have seen that, assuming ZDSAF or DF relaying protocols, the achievable rates are not necessarily concave.

Assuming that the transmitters can be coordinated, and by using the time-sharing scheme similarly to Subsec. III-D, the achievable transmission rate of user $i$ is given by:

$$u_i^{\mathrm{TS}}(\underline{\theta}_i, \underline{\theta}_{-i}) = \sum_{q=1}^{Q} R_i^{(q),\mathrm{TS}}(\theta_i^{(q)}, \theta_{-i}^{(q)}) \tag{24}$$

where $R_i^{(q),\mathrm{TS}}$ is the rate user $i$ obtains by using band (q) and time-sharing technique. After Sec. III-D the latter quantity is:

$$\forall i \in \{1,2\}, R_i^{(q),\mathrm{TS}} \;=\; \alpha_i^{(q)}\overline{\beta_j}^{(q)} R_i^{(q)}\left(\frac{\theta_i^{(q)}P_i}{\alpha_i}, 0\right) + \alpha_i^{(q)}\beta_j^{(q)} R_i^{(q)}\left(\frac{\theta_i^{(q)}P_i}{\alpha_i}, \frac{\theta_j^{(q)}P_j}{\alpha_j}\right), \tag{25}$$

where $j = -i$, $(\alpha_i^{(q)}, \alpha_j^{(q)}) \in [0,1]^2$, $(\beta_i^{(q)}, \beta_j^{(q)}) \in [0,1]^2$ such that $\beta_1^{(q)}\alpha_2^{(q)} = \beta_2^{(q)}\alpha_1^{(q)}$. These parameters are fixed and chosen such that the achievable rates are maximized. The rates $R_i^{(q)}\left(\frac{\theta_i^{(q)}P_i}{\alpha_i}, 0\right)$, $R_i^{(q)}\left(\frac{\theta_i^{(q)}P_i}{\alpha_i}, \frac{\theta_j^{(q)}P_j}{\alpha_j}\right)$ represent the achievable rates in band $(q)$ when time-sharing is used. These rates depend on the relaying protocol and are given by Eq. (13) for DF and by Eq. (20) for ZDSAF. Notice that, when EF is assumed, the rates are always concave irrespective of the channel gains and time-sharing techniques do not change the achievable rate-region.

*Theorem 4.6:* [Existence of an NE for TS] *There always exists at least one pure NE in the PA game $\mathcal{G}^{\mathrm{TS}}$, regardless of the used relaying scheme and the values of the channel gains.*

If the users are coordinated (i.e., each user is aware of the moments where the other user is transmitting or not) then their achievable rates $R_i^{(q),\mathrm{TS}}$ are always concave w.r.t. $\theta_i^{(q)}$. This implies directly that [27], irrespective of the relaying technique and of the channel gains, the existence of an NE will be guaranteed.

In the particular case where either $P_r^{(q)} = 0$ or $h_{ir}^{(q)} = 0$, for all $q \in \{1, \dots, Q\}$ and $i \in \{1, 2\}$, the parallel IRC reduces to the parallel interference channel [6]. The time-sharing scheme is useless since the achievable rates are already concave and $\alpha_i = 1$, $\beta_i = 1$ are optimal. Therefore, Theorem 4.6 guarantees the existence of the NE in this case and is consistent with the known results in [6].

## V. Simulation results

*Single-band IRCs: AF vs DF vs EF.* Here, we assume $Q = 1$ and a path loss exponent of $2$ that is, $|h_{ij}| = \left(\frac{d_{ij}}{d_0}\right)^{-\frac{\gamma}{2}}$ for $(i,j) \in \{1, 2, r\}^2$ where $d_0 = 5$ m is a reference distance and $\gamma = 2$ is the path loss exponent. The nodes $\mathcal{S}_1$, $\mathcal{S}_2$, $\mathcal{D}_1$, $\mathcal{D}_2$ are assumed to be located in a plane. The positions of the nodes will be indicated on each figure and are characterized by the distance between them which are chosen as follows: $d'_{11} = 11.5$ m, $d'_{22} = 10$ m, $d'_{12} = 11$ m and $d'_{21} = 14$ m. As for the relay, to avoid any divergence for the path loss in $d_{ij} = 0$, we assume that it is located a hight $\epsilon = 0.1$ m from this plane i.e., the relay location is given by the $(x_r, y_r, z_r)$ where $z_r$ is fixed and equals $0.1$ m; thus $d_{ij} = \sqrt{d'^2_{ij} + \epsilon^2}$ for $i = r$ or $j = r$ and $i \neq j$. The noise levels at the receiver nodes are assumed to be normalized ($N_1 = N_2 = N_r = 1$). In terms of transmit power we analyze two cases: a symmetric case where $P_1 = P_2 = 10$ (normalized power) and an asymmetric one where $P_1 = 3$ and $P_2 = 10$. The relay transmit power is fixed: $P_r = 10$. For the symmetric scenario, Fig. 2 represents the regions of the plane $\left(\frac{x_r}{d_0}, \frac{y_r}{d_0}\right) \in [-4, +4] \times [-3, +4]$ (corresponding to the possible relay positions) where a certain protocol performs better than the two others in terms of system sum-rate. These regions are in agreement with what is generally observed for the standard relay channel.

This type of information is useful, for example, when the relay has to be located in specific places because of different practical constraints and one has to choose the best protocol. Fig. 3 allows one to better quantify the differences in terms of sum-rate between the AF, DF and bi-level EF protocols since it represents the sum-rate versus $x_r$ for a given $y_r = 0.5d_0$. The discontinuity observed stems from the fact that for the bi-level EF protocol there is a frontier delineating the scenarios where one receiver is better than the other and can therefore suppress the interference of the relay (as explained in Sec. III-B).

*Number of Nash equilibria for the AF protocol.* First, we show that in the PA game with ZDSAF, one can have three possible Nash equilibria. For a given typical scenario composed of an IC in parallel with an IRC ($Q = 2$) and $\rho_1 = 1$, $\rho_2 = 3$, $\rho_r = 2$ and the channel gains $(g_{11}, g_{12}, g_{21}, g_{22}) = (2.76, 5.64, -3.55, -1.61)$, $(h_{11}, h_{12}, h_{21}, h_{22}) = (14.15, 3.4, 0, 1.38)$ and $(h_{1r}, h_{2r}, h_{r1}, h_{r2}) = (-3.1, 2.22, -3.12, 1.16)$, we plot the best response functions in Fig. 4. We see that there are three intersection points and therefore three Nash equilibria. As explained in Sec. IV-C, the effectively observed NE in a one-shot game is not predictable without any additional assumptions. However, the Cournot tatônnement procedure converges towards a given NE which can be predicted from the sole knowledge of the starting point of the game, namely $\theta_1^0$ or $\theta_2^0$.

*Stackelberg formulation.* We have mentioned that a strong motivation for studying IRCs is to be able to introduce relays in a network with non-coordinated and interfering pairs of terminals. For example, relays could be introduced by an operator aiming at improving the performance of the communications of his customers. In such a scenario, the operator acts as a player and more precisely as a game leader in the sense of [35]. In [35], the author introduced what is called nowadays a Stackelberg game. This type of hierarchical games comprises one leader which plays in the first step of the game and several players (the followers) which observe the leader's strategy and choose their actions accordingly. In our context, the game leader is the operator/engineer /relay who chooses the parameters of the relays. The followers are the adaptive/cognitive transmitters that adapt their PA policy to what they observe. In the preceding sections we have mentioned some of these parameters: the location of each relay; in the case of AF, the amplification gain of each relay; in the case of DF and EF, the power allocation policy between the two cooperative signals at each relay i.e., the parameter $\nu^{(q)}$. Therefore, the relay can be thought of as a player who maximizes its own utility. This utility can be either the individual utility of a given transmitter (picture one WiFi subscriber wanting to increase his downlink throughput by locating his cellular phone somewhere in his apartment while his neighbor can also exploit the same spectral resources) or the network sum-rate (in the case of an operator). In the latter case, the operator possesses some degrees of freedom to improve the efficiency of the equilibrium. In the remaining part of this section, we focus on the Stackelberg formulation where the strategy of the leader is respectively the relay amplification factor, position and power allocation between the cooperative signals. The system considered is composed of an IRC in parallel with an interference channel (IC) [36]. All the simulations provided are obtained by applying the Cournot tatônnement procedure. The simulation setup is as follows. The source and destination nodes are located in fixed locations in the region $[-L, L]^2$ of a plane, with $L = 10$ m, such that the relative distances between the nodes are: $d_{11} = 6.52\text{m}$, $d_{12} = 8.32\text{m}$, $d_{21} = 6.64\text{m}$, $d_{22} = 6.73\text{m}$. We assume a path loss model for the channel gains $|g_{ij}|$, $|h_{ij}|$. For the path loss model we take $|h_{ij}| = \left(\frac{d_{ij}}{d_0}\right)^{-\frac{\gamma^{(1)}}{2}}$ and $|g_{ij}| = \left(\frac{d_{ij}}{d_0}\right)^{-\frac{\gamma^{(2)}}{2}}$ for $(i, j) \in \{1, 2, r\}^2$ where $d_0 = 1$ m is a reference distance. The relay is at $\epsilon = 0.5$ m from the plane. We will also assume that $N_i^{(1)} = N_i^{(2)} = N_i$, $i \in \{1, 2\}$, $N_r^{(1)}$ will be denoted by $N_r$ and also $P_r^{(1)} = P_r$, $A_r^{(1)} = A_r$, $a_r^{(1)} = a_r$, $\tilde{a}_r^{(1)} = a_r$, $\nu^{(1)} = \nu$.

*Optimal relay amplification gain for the AF protocol.* First we consider the ZDSAF relaying scheme assuming a fixed amplification gain $a_r = A_r$ (Sec. IV-C). We want to analyze the influence of the value of the amplification factor, $A_r \in [0, \tilde{a}_r(1, 1)]$, on the achievable network sum-rate at the NE. This is what Fig. 5 shows for the following scenario: $\epsilon = 0.5$ m, $P_1 = 20$ dBm, $P_2 = 23$ dBm, $P_r = 22$ dBm, $N_1 = 10$ dBm, $N_2 = 9$ dBm, $N_r = 7$ dBm, $\gamma^{(1)} = \gamma^{(2)} = 2$. We observe that the optimal value is $A_r^* = 0.05$ and is not equal to the one saturating the relay power constraint $\tilde{a}_r(1, 1) = 0.17$. This result illustrates for the sum-rate what we have proved analytically for the individual rate of a given user (see Sec. III-C). Note that the gap between the optimal choice for $a_r$ and the choice saturating the power constraint is not that large and in fact other simulation results have shown is is generally of this order and even smaller typically.

*Optimal relay location for the AF protocol.* Now, we consider the ZDSAF when the full power regime is assumed at the relay, $a_r = \tilde{a}_r(\theta_1, \theta_2)$ (Sec. IV-C) and study the relay location problem. Fig. 6 represents the achievable network sum-rate as a function of the relay position $(x_{\mathcal{R}}, y_{\mathcal{R}}) \in [-L, L]^2$ for the scenario: $P_1 = 20$ dBm, $P_2 = 17$ dBm, $P_r = 22$ dBm, $N_1 = 10$ dBm, $N_2 = 9$ dBm, $N_r = 7$ dBm, $\gamma^{(1)} = 2.5$ and $\gamma^{(2)} = 2$. We observe that there are two local maximum that actually correspond to the points that maximize the individual achievable rates. Many simulation results have confirmed that, when the source nodes are sufficiently far away from each other, maximizing the individual rate of either user at the NE amounts to locating the relay on one of the the segment between $\mathcal{S}_i$ and $\mathcal{D}_i$. This interesting and quite generic observation can be explained as follows. For this purpose, consider Fig. 6 which is a temperature image representing the values of $\theta_1$ and $\theta_2$ for different relay positions in $[-L, L]^2$. The region where $(\theta_1, \theta_2) = (1, 0)$ (resp. $(\theta_1, \theta_2) = (0, 1)$) is the region around $\mathcal{S}_1$ (resp. $\mathcal{S}_2$). We see that the intersection between these regions corresponds to a small area. This quite general observation shows that the selfish behavior of the transmitters leads to self-regulating the interference in the network. Said otherwise, a selfish transmitter will not use at all a far away relay but leaves it to the other transmitter. Thus, when one transmitter uses the relay, it is often alone and sees no interference. In these conditions, by considering the path loss effects it can be proved that the optimal relay position is on the segment between the considered source and destination nodes. This also explains why the position that maximizes the network sum-rate lies also on one of the segments from $\mathcal{S}_i$ to $\mathcal{D}_i$.

*Optimal relay power allocation at the relay for DF and EF.* For the DF protocol, we fix the cooperation degrees $\tau_1 = 0$ and $\tau_2 = 0$. In Fig. 7, we plot the achievable sum-rate at the equilibrium as a function of the relay power allocation policy is $\nu \in [0, 1]$ (with the convention $\nu = \nu^{(1)}$) for the scenario: $x_{\mathcal{R}} = 0$ m, $y_{\mathcal{R}} = 0$ $m$, $P_1 = 22$ dBm, $P_2 = 17$ dBm, $P_r = 23$ dBm, $N_1 = 7$ dBm, $N_2 = 9$ dBm, $N_r = 0$ dBm, $\gamma^{(1)} = 2.5$ and $\gamma^{(2)} = 2$. We observe that, for both protocols, the optimal power allocation $\nu^* = 1$, meaning that the relay allocates all its available power to the better receiver, $\mathcal{D}_1$. In this case, the relay is in very good conditions and can therefore reliably decode the source messages. This explains why DF outperforms EF which is in agreement with the observations we have made in Sec. III. We have observed that, in general, the network sum-rate is not concave w.r.t. $\nu \in [0, 1]$ and that the optimal power allocation lies on the borders $\nu^* \in \{0, 1\}$ for both relaying protocols. In Fig. 7, we also see that the fair PA policy that is, $\nu = \frac{1}{2}$ can lead to a relatively significant performance loss.

## VI. CONCLUSION

The complete study of PA games in IRCs is a wide problem and we do not claim to fully characterize it here. One of the main objectives in this paper has been to know whether there exist some stable outcomes to the conflicting situation where two transmitters selfishly allocate their power between different sub-channels in multi-band interference relay channels in order to maximize their individual transmission rate. Our approach has been to consider transmission rates achievable in a decentralized framework where relays can be deployed with minor or even with no changes for the already existing receivers. For the three types of protocols considered, we have proved that the utility of the transmitters is a concave function of the individual strategy, which ensures the existence of Nash equilibria in the power allocation game after Rosen [27]. In a special case of the AF protocol, we have fully characterized the number of NE and the convergence problem of Cournot-type or iterative water-filling procedures to an NE. Although we have limited the scope of the paper, we have seen that studying IRCs deeply requires further investigations. Many interesting questions which can be considered as natural extensions of this work have arisen. Considering more efficient coding-decoding schemes and relaying protocols such as those of [15] and related works, is it possible to prove that the utilities are still concave functions? For these schemes and those considered in this paper, it is also important to fully determine the number of Nash equilibria and derive convergent iterative distributed power allocation algorithms. We have also seen that several power allocation games come into play and need to be studied when considering DF, EF and AF-type protocols: for allocating transmit power between the different bands at the sources, for choosing the cooperation degree at the sources, for allocating the power between the cooperation signals at the relay, for allocating the transmit power over time. Furthermore, a new agent can come into play (the relay) and several Stackelberg formulations can be used to improve the efficiency of the equilibria.

**Acknowledgments**

The authors would like to thank Prof. Pierre Duhamel, Prof. Jean-Claude Belfiore, and Prof. Luc Vandendorpe for their useful feedbacks on some parts of this work.

## Appendix A

### Proof of Theorem 3.2 (achievable transmission rates for IRCs with the EF protocol)

In order to prove that the transmission rate region of Theorem 3.2 is achievable for Gaussian IRCs, we use a quite common approach [20] for proving coding theorems: we first prove that it is achievable for discrete input discrete output channels and obtain the Gaussian case from standard quantization and continuity arguments [20], and a proper choice of coding auxiliary variables.

**Definitions and notations**

We denote by $A_\epsilon^{(n)}(X)$ the weakly $\epsilon$-typical set for the random variable $X$. If $X$ is a discrete variable, $X \in \mathcal{X}$, then $\|\mathcal{X}\|$ denotes the cardinality of the finite set $\mathcal{X}$. We use $x^n$ to indicate the vector $(x_1, x_2, \ldots, x_n)$.

*Definition A.1: A two-user discrete memoryless interference relay channel (DMIRC) without feedback consists of three input alphabets $\mathcal{X}_1$, $\mathcal{X}_2$ and $\mathcal{X}_r$, and three output alphabets $\mathcal{Y}_1$, $\mathcal{Y}_2$ and $\mathcal{Y}_r$, and a probability transition function that satisfies $p\left(y_1^n, y_2^n, y_r^n \mid x_1^n, x_2^n, x_r^n\right) = \prod_{k=1}^{n} p\left(y_{1,k}, y_{2,k}, y_{r,k} \mid x_{1,k}, x_{2,k}, x_{r,k}\right)$ for some $n \in \mathbb{N}^*$.*

*Definition A.2: A $\left(2^{nR_1}, 2^{nR_2}, n\right)$-code for the DMIRC with private messages consists of two sets of*

*integers* $\mathcal{W}_1 = \left\{1,...,2^{nR_1}\right\}$ *and* $\mathcal{W}_2 = \left\{1,...,2^{nR_2}\right\}$, *two encoders:* $f_i : \mathcal{W}_i \to \mathcal{X}_i^n$,, *a set of relay functions* $\{f_{r,k}\}_{k=1}^n$ *such that* $x_{r,k} = f_{r,k}\left(y_{r,1}, y_{r,2}, ..., y_{r,k-1}\right),\ 1 \le k \le n$ *and two decoding functions* $g_i : \mathcal{Y}_i^n \to \mathcal{W}_i,\ i \in \{1,2\}$. *The source node* $\mathcal{S}_i$ *intends to transmit* $W_i$, *the private message, to the receiver node* $\mathcal{D}_i$.

*Definition A.3: The average probability of error is defined as the probability that the decoded message pair differs from the transmitted message pair; that is,* $P_e^{(n)} = \Pr\left[g_1\left(Y_1^n\right) \neq W_1 \text{ or } g_2\left(Y_2^n\right) \neq W_2 \mid (W_1, W_2)\ \right]$, *where* $(W_1, W_2)$ *is assumed to be uniformly distributed over* $\mathcal{W}_1 \times \mathcal{W}_2$. *We also define the the average probability of error for each receiver as* $P_{ei}^{(n)} = \Pr\left[g_i\left(Y_i^n\right) \neq W_i \mid W_i\right]$. *We have* $0 \le \max\left\{P_{e1}^{(n)}, P_{e2}^{(n)}\right\} \le P_e^{(n)} \le P_{e1}^{(n)} + P_{e2}^{(n)}$. *Hence* $P_e^{(n)} \to 0$ *implies that both* $P_{e1}^{(n)} \to 0$ *and* $P_{e2}^{(n)} \to 0$, *and conversely.*

*Definition A.4: A rate pair* $(R_1, R_2)$ *is said to be achievable for the IRC if there exists a sequence of* $\left(2^{nR_1}, 2^{nR_2}, n\right)$ *codes with* $P_e^{(n)} \to 0$ *as* $n \to \infty$.

**Overview of coding strategy**

At the end of the block $k$, the relay constructs two estimations $\hat{y}_{r1}^n(k)$ and $\hat{y}_{r2}^n(k)$ of its observation $y_r^n(i)$ that intends to transmit to the receivers $\mathcal{D}_1$ and $\mathcal{D}_2$ to help them resolve the uncertainty on $w_{1,k}$ and $w_{2,k}$ respectively at the end of the block $k+1$.

**Details of the coding strategy**

Codebook generation

- i Generate $2^{nR_i}$ i.i.d. codewords $x_i^n(w_i) \sim \prod_{k=1}^n p\left(x_{i,k}\right)$, where $w_i \in \left\{1, \ldots, 2^{nR_i}\right\}$, $i \in \{1,2\}$.
- ii Generate $2^{nR_0^{(1)}}$ i.i.d. codewords $u_1^n \sim \prod_{k=1}^n p\left(u_{1,k}\right)$. Label these $u_1^n(s_1),\ s_1 \in \left\{1, \ldots, 2^{nR_0^{(1)}}\right\}$.
- iii Generate $2^{nR_0^{(2)}}$ i.i.d. codewords $u_2^n \sim \prod_{k=1}^n p\left(u_{2,k}\right)$. Label these $u_2^n(s_2),\ s_1 \in \left\{1, \ldots, 2^{nR_0^{(2)}}\right\}$.
- iv For each pair $(u_1^n(s_1), u_2^n(s_2))$, choose a sequence $x_r^n$ where $x_r^n \sim p\left(x_r^n | u_1^n(s_1), u_2^n(s_2)\right) = \prod_{k=1}^n p\left(x_{r,k} | u_{1,k}(s_1), u_{2,k}(s_2)\right)$.
- v For each $u_1^n(s_1)$, generate $2^{n\hat{R}_1}$ conditionally i.i.d. codewords $\hat{y}_{r1}^n \sim \prod_{k=1}^n p\left(\hat{y}_{r1k} | u_{1,k}(s_1)\right)$ and label them $\hat{y}_{r1}^n(z_1|s_1),\ z_1 \in \left\{1, \ldots, 2^{n\hat{R}_1}\right\}$. For each pair $(u_1, \hat{y}_{r1}) \in \mathcal{U}_1 \times \hat{\mathcal{Y}}_{r1}$, the conditional probability $p(\hat{y}_{r1}|u_1)$ is defined as $p(\hat{y}_{r1}|u_1) = \sum_{x_1,x_2,y_1,y_2,y_r} p\left(x_1\right) p\left(x_2\right) p\left(y_1, y_2, y_r | x_1, x_2, x_r\right) p\left(\hat{y}_{r1} | y_r, u_1\right)$.
- vi For each $u_2^n(s_2)$, generate $2^{n\hat{R}_2}$ conditionally i.i.d. codewords $\hat{y}_{r2}^n \sim \prod_{k=1}^n p\left(\hat{y}_{r2k} | u_{2,k}(s_2)\right)$ and label them $\hat{y}_{r2}^n(z_2|s_2),\ z_2 \in \left\{1, 2^{n\hat{R}_2}\right\}$. For each triplet $(u_2, \hat{y}_{r1}) \in \mathcal{U}_2 \times \hat{\mathcal{Y}}_{r1}$, the conditional probability $p(\hat{y}_{r2}|u_2)$ is defined as $p(\hat{y}_{r2}|u_2) = \sum_{x_1,x_2,y_1,y_2,y_r} p\left(x_1\right) p\left(x_2\right) p\left(y_1, y_2, y_r | x_1, x_2\right) p\left(\hat{y}_{r2} | y_r, u_2\right)$.
- vii Randomly partition the message set $\left\{1, 2, \ldots, 2^{n\hat{R}_1}\right\}$ into $2^{nR_0^{(1)}}$ sets $\left\{S_1^{(1)}, S_2^{(1)}, \ldots, S_{2^{nR_0^{(1)}}}^{(1)}\right\}$ by independently and uniformly assigning each message in $\left\{1,\ldots,2^{n\hat{R}_1}\right\}$ to an index in $\left\{1,\ldots,2^{nR_0^{(1)}}\right\}$.

viii Also, randomly partition the message set $\left\{1,2,\ldots,2^{n\hat{R}_2}\right\}$ into $2^{nR_0^{(2)}}$ sets $\left\{S_1^{(2)},S_2^{(2)},\ldots,S_{2^{nR_0^{(2)}}}^{(2)}\right\}$ by independently and uniformly assigning each message in $\left\{1,\ldots,2^{n\hat{R}_2}\right\}$ to an index in $\left\{1,\ldots,2^{nR_0^{(2)}}\right\}$.

Encoding procedure Let $w_{1,k}$ and $w_{2,k}$ be the messages to be send in block $k$. $\mathcal{S}_1$ and $\mathcal{S}_2$ respectively transmit the codewords $x_1^n(w_{1,k})$ and $x_2^n(w_{2,k})$. We assume that $(u_1^n(s_{1,k-1}),\hat{y}_{r1}^n(z_{1,k-1}|s_{1,k-1}),y_r^n(k-1))\in A_\epsilon^{(n)}$ and $z_{1,k-1}\in S_{s_{1,k}}^{(1)}$ and also that $(u_2^n(s_{2,k-1}),\hat{y}_{r2}^n(z_{2,k-1}|s_{2,k-1}),y_r^n(k-1))\in A_\epsilon^{(n)}$ with $z_{2,k-1}\in S_{s_{2,k}}^{(2)}$. Then the relay transmits the codeword $x_r^n(s_{1,k},s_{2,k})$.

Decoding procedure In what follows, we will only detail the decoding procedure at the receiver node $\mathcal{D}_1$ (at $\mathcal{D}_2$ the decoding is analogous). At the end of block $k$:

i The receiver node $\mathcal{D}_1$ estimates $\hat{s}_{1,k}=s_1$ if and only if there exists a unique sequence $u_1^n(s_1)$ that is jointly typical with $y_1^n(k)$. We have $s_1=s_{1,k}$ with arbitrarily low probability of error if $n$ is sufficiently large and $R_0^{(1)}<I(U_1;Y_1)$.

ii Next, the receiver node $\mathcal{D}_1$ constructs a set $\mathcal{L}_1\left(y_1^n\left(k-1\right)\right)$ of indexes $z_1$ such that $\left(u_1^n\left(\hat{s}_{1,k-1}\right),\hat{y}_{r1}^n\left(z_1|\hat{s}_{1,k-1}\right),y_1^n\left(k-1\right)\right)\in A_\epsilon^{(n)}$. $\mathcal{D}_1$ estimates $\hat{z}_{1,k-1}$ by doing the intersection of sets $\mathcal{L}_1\left(y_1^n\left(k-1\right)\right)$ and $S_{\hat{s}_{1,k}}^{(1)}$. Similarly to [14, Theorem 6] and using [14, Lemma 3], one can show that $\hat{z}_{1,k-1}=z_{1,k-1}$ with arbitrarily low probability of error if $n$ is sufficiently large and $\hat{R}_1<I(\hat{Y}_{r1};Y_1|U_1)+R_0^{(1)}$.

iii Using $\hat{y}_{r1}^n(\hat{z}_{1,k-1}|\hat{s}_{1,k-1})$ and $y_1^n(k-1)$, the receiver node $\mathcal{D}_1$ finally estimates the message $\hat{w}_{1,k-1}=w_1$ if and only if there exists a unique codeword $x_1^n(w_1)$ such that $(x_1^n(w_1),u_1^n(\hat{s}_{1,k-1}),y_1^n(i-1),\hat{y}_{r1}^n(\hat{z}_{1,k-1}|\hat{s}_{1,k-1}))\in A_\epsilon^{(n)}$. One can show that $w_1=w_{1,k-1}$ with arbitrarily low probability of error if $n$ is sufficiently large and

$$R_1 \;<\; I\left(X_1;Y_1,\hat{Y}_{r1}\mid U_1\right). \tag{26}$$

iv At the end of the block $k$, the relay looks for the suitable estimation of its observation that it intends to transmit to the receiver node $\mathcal{D}_1$ by estimating $\hat{z}_{1,k}$. It estimates $\hat{z}_{1,k}=z_1$ if there exists a sequence $\hat{y}_r^n(z_1|s_{1,k})$ such that $(u_1^n(s_{1,k}),\hat{y}_{r1}^n(z_1|s_{1,k}),y_r^n(k))\in A_\epsilon^{(n)}$. There exists a such sequence if $n$ is sufficiently large and $\hat{R}_1>I(\hat{Y}_{r1};Y_r|U_1)$.

From i, ii, iii we further obtain

$$I(\hat{Y}_{r1};Y_r|U_1,Y_1) \;<\; I(U_1;Y_1). \tag{27}$$

$$R_i^{'}(a_r) = \frac{a_r^2\left[|m_i|^2\mathrm{Re}(p_iq_i^*) - (|p_i|^2+s_i)\mathrm{Re}(m_in_i^*)\right] + a_r\left[|m_i|^2(|q_i|^2+1) - |n_i|^2(|p_i|^2+s_i)\right] + (|q_i|^2+1)\mathrm{Re}(m_in_i^*) - |n_i|^2\mathrm{Re}(p_iq_i^*)}{[|p_ia_r+q_i|^2+s_ia_r^2+1][|m_ia_r+n_i|^2+|p_ia_r+q_i|^2+s_ia_r^2+1]} \quad (28)$$

The achievability proof for the second receiver node follows in a similar manner. Therefore, we have completed the proof.

**From the discrete case to the Gaussian case**

As mentioned in the beginning of this section, obtaining achievable transmission rates for Gaussian IRCs from those for discrete IRCs is an easy task. Indeed, the latter consists in using Gaussian codebooks everywhere and choosing the coding auxiliary variables properly namely choosing $U_1$, $U_2$, $\hat{Y}_{r,1}$, and $\hat{Y}_{r,2}$. The coding auxiliary variables $U_1$ and $U_2$ are chosen to be independent and distributed as $U_1 \sim \mathcal{N}(0, \nu_1 P_r)$ and $U_2 \sim \mathcal{N}(0, \nu_2 P_r)$. The corresponding codewords $u_1^n$ and $u_2^n$ convey the messages resulting from the compression of $Y_r$. The auxiliary variables $\hat{Y}_{r,1}$, $\hat{Y}_{r,2}$ write as $\hat{Y}_{r,1} = Y_r + Z_{wz}^{(1)}$ and $\hat{Y}_{r,2} = Y_r + Z_{wz}^{(2)}$ where the compression noises $Z_{wz}^{(1)} \sim \mathcal{N}(0, N_{wz}^{(1)})$ and $Z_{wz}^{(2)} \sim \mathcal{N}(0, N_{wz}^{(2)})$ are independent. At last, the relay transmits the signal $X_r = U_1 + U_2$ as in the case of a broadcast channel except that, here, each destination also receives two direct signals from the source nodes. By making these choices of random variables we obtain the desired rate region.

## APPENDIX B
## PROOF OF THEOREM 3.4 (OPTIMAL AMPLIFICATION GAIN FOR ZDSAF IN IRCS)

Using the notations given in Theorem 3.4 and also the signal-to-noise plus interference ratio in the capacity function of Eq. (10) the rate $R_i$ can be written as:

$$R_i(a_r) = C\left(\frac{|m_ia_r+n_i|^2}{|p_ia_r+q_i|^2+s_ia_r^2+1}\right).$$

We observe that $R_i(0) = C\left(\frac{|n_i|^2}{|q_i|^2+1}\right)$ and that we have an horizontal asymptote $R_{i,\infty} \triangleq \lim_{a_r\to\infty} R_1(a_r) = C\left(\frac{|m_i|^2}{|p_i|^2+s_i}\right)$. Also the first derivative w.r.t. $a_r$ is$a_r$ is given in (28)

The explicit solution, $a_r^*$ depends on the channel parameters and is given here below. We denote by $\Delta$ the discriminant of the nominator in the previous equation. If $\Delta < 0$, then in function of the sign of $|m_i|^2\mathrm{Re}(p_iq_i^*) - (|p_i|^2+s_i)\mathrm{Re}(m_in_i^*)$, the function $R_i(a_r)$ is either decreasing and $a_r^* = 0$ or increasing and $a_r^* = \overline{a}_r$. Let us now focus on the case where $\Delta \geq 0$.

1) If $|m_i|^2\mathrm{Re}(p_i q_i^*) - (|p_i|^2 + s_i)\mathrm{Re}(m_i n_i^*) \geq 0$ then
   a) if $c_{r,i}^{(1)} \leq 0$ and $c_{r,i}^{(2)} \leq 0$ then $a_r^* = \overline{a}_r$;
   b) if $c_{r,i}^{(1)} > 0$ and $c_{r,i}^{(2)} \leq 0$ then
      i) if $\overline{a}_r \geq c_{r,i}^{(1)}$ then $a_r^* = 0$;
      ii) if $\overline{a}_r < c_{r,i}^{(1)}$ then
         - if $R_i(0) \geq R_i(\overline{a}_r)$ then $a_r^* = 0$ else $a_r^* = \overline{a}_r$;
   c) if $c_{r,i}^{(1)} \leq 0$ and $c_{r,i}^{(2)} > 0$ then the analysis is similar to the previous case and $a_r^* \in \{0, \overline{a}_r\}$ depending on $a_r^{(2)}$ this time;
   d) if $c_{r,i}^{(1)} > 0$ and $c_{r,i}^{(2)} > 0$
      i) if $c_{r,i}^{(1)} < c_{r,i}^{(2)}$
         A) if $\overline{a}_r \leq c_{r,i}^{(1)}$ then $a_r^* = \overline{a}_r$;
         B) if $c_{r,i}^{(1)} < \overline{a}_r \leq c_{r,i}^{(2)}$ then $a_r^* = c_{r,i}^{(1)}$;
         C) if $\overline{a}_r > c_{r,i}^{(2)}$ then
            - if $R_i(c_{r,i}^{(1)}) \geq R_1(\overline{a}_r)$ then $a_r^* = c_{r,i}^{(1)}$ else $a_r^* = \overline{a}_r$;
      ii) if $c_{r,i}^{(1)} > c_{r,i}^{(2)}$ then the analysis is similar to the previous case, exchanging the roles of $c_{r,i}^{(1)}$ and $c_{r,i}^{(2)}$;
      iii) if $c_{r,i}^{(1)} = c_{r,i}^{(2)}$ then $a_r^* = \overline{a}_r$.
2) If $|m_i|^2\mathrm{Re}(p_i q_i^*) - (|p_i|^2 + s_i)\mathrm{Re}(m_i n_i^*) < 0$ then the analysis follows in the same lines and $a_r^* \in \{0, \overline{a_r}, c_{r,i}^{(1)}, c_{r,i}^{(2)}\}$.

## APPENDIX C

### PROOF OF THEOREM 4.3 (EXISTENCE OF AN NE FOR THE BI-LEVEL COMPRESSION EF PROTOCOL)

We want to prove that for each user $R_i^{(q)}$ is concave w.r.t. $\theta_i^{(q)}$. Consider w.l.o.g. the case of user 1. The general case of complex channel gains is considered. We analyze the second derivative of $R_1^{(q)}$ given in Eq. (17). For the sake of clarity we denote by $\tilde{N}_1^{(q)} = |h_{r1}|^2\overline{\nu^{(q)}}P_r^{(q)} + N_1^{(q)}$, $\Gamma_0 = |h_{r1}|^2\nu^{(q)}P_r^{(q)}$ and $\Gamma_1 = |h_{21}|^2\theta_2^{(q)}P_2 + \tilde{N}_1^{(q)}$. After some manipulations we obtain the following relation: $\frac{d^2R_1^{(q)}}{d(\theta_1^{(q)})^2} = M_1 - M_2$ with $M_k = \frac{NM_k}{DM_k}$, $k \in \{1, 2\}$ where (for the sake of clarity we have denoted $h_{ij}^{(q)}$ by $h_{ij}$):$NM_1$, $NM_2$, $DM_1$, $DM_2$ are defined by (29),(30).

We observe that the terms $\Lambda_k \geq 0$, $k \in \{2, \ldots, 7\}$. Also we can easily see from Eq. (29) that $M_2 \geq 0$, $DM_1 \geq 0$. Thus if we prove that $NM_1 \leq 0$ the concavity of $R_1^{(q)}$ will be guaranteed. In this purpose we

$$\begin{cases} NM_1 &= 2\frac{\left(|h_{11}|^2 {P_1}^2 |h_{1r}|^2 - {\Lambda_2}^2 {P_1}^2\right)|h_{11}|^2 \theta_1^{(q)}\, P_1}{\Gamma_0\, \Lambda_5} + 2\,\frac{\Lambda_8\, |h_{11}|^2 P_1}{\Gamma_0\, \Lambda_5} - \\ & 2\left(\frac{\Lambda_8\, |h_{11}|^2 \theta_1^{(q)}\, P_1}{\Gamma_0} + \Lambda_6\, |h_{11}|^2 P_1 + |h_{1r}|^2\, \Gamma_1\, P_1\right)\frac{\Lambda_8\, \Gamma_1}{{\Lambda_5}^2 \Gamma_0} \\ & +2\,\frac{\Lambda_7\, {\Lambda_8}^2 {\Gamma_1}^2}{{\Gamma_0}^2 {\Lambda_5}^3} - 2\,\frac{\Lambda_7\left(|h_{11}|^2 {P_1}^2 |h_{1r}|^2 - {\Lambda_2}^2 {P_1}^2\right)\Gamma_1}{{\Lambda_5}^2 \Gamma_0}, \\ NM_2 &= \left[\left(\frac{\Lambda_8\, |h_{11}|^2 \theta_1^{(q)}\, P_1}{\Gamma_0} + \Lambda_6\, |h_{11}|^2 P_1 + |h_{1r}|^2\, \Gamma_1\, P_1\right)\frac{1}{\Lambda_5} - \frac{\Lambda_7\, \Lambda_8\, \Gamma_1}{{\Lambda_5}^2 \Gamma_0}\right]^2, \\ DM_1 &= 1 + \frac{\Lambda_7}{\Lambda_5}, \\ DM_2 &= DM_1^2, \end{cases} \tag{29}$$

$$\begin{cases} \Lambda_1 &= 2\mathcal{R}e(h_{11} h_{1r}^* h_{21}^* h_{2r}), \\ \Lambda_2 &= \left|h_{11} h_{1r}^*\right|, \\ \Lambda_3 &= A^{(q)} \\ \Lambda_4 &= |h_{11}|^2\, \theta_1^{(q)} P_1 + \Gamma_1, \\ \Lambda_5 &= \left(N_r^{(q)} + N_{wz,1}^{(q)}\right)\Gamma_1 + |h_{2r}|^2\, \theta_2^{(q)} P_2 \tilde{N}_1^{(q)}, \\ \Lambda_6 &= |h_{2r}|^2\, \theta_2^{(q)} P_2 + N_r^{(q)} + N_{wz,1}^{(q)}, \\ \Lambda_7 &= \Lambda_6\, |h_{11}|^2\, \theta_1^{(q)} P_1 + \Gamma_1\, |h_{1r}|^2\, \theta_1^{(q)} P_1, \\ \Lambda_8 &= |h_{11}|^2\, P_1 \Lambda_3 + |h_{1r}|^2\, P_1 \Lambda_4 - 2\Lambda_2^2 \theta_1^{(q)} P_1^2 - \Lambda_1 \theta_2^{(q)} P_1 P_2. \end{cases} \tag{30}$$

plug the expressions of $\Lambda_5$, $\Lambda_6$, $\Lambda_7$, $\Lambda_8$ into Eq. (29) and obtain that $NM_1 = \frac{NNM_1}{DNM_1}$ with $NNM_1$, $DNM_1$ given in (31), (32).

Therefore we obtain the desired result $NM_1 \leq 0$ and thus $M_1 \geq 0$, which implies that $\frac{d^2 R_1^{(q)}}{d\left(\theta_1^{(q)}\right)^2} \leq 0$, whatever the channel parameters.

## Appendix D

## Proof of Theorem 4.5 (Number of Nash equilibria for ZDSAF)

Before discussing these situations in detail, let us first observe that the two functions $F_i(\theta_j)$ are decreasing w.r.t. $\theta_j$ and also $F_i(0) = \frac{d_i}{c_{ii}}$, $F_i(\theta_j^o) = 0$ where $\theta_j^o = \frac{d_i}{c_{ij}}$.

In this section, we will investigate the NE of the game and also their asymptotical stability of each NE point. A sufficient and necessary condition that guarantees the asymptotic stability of a certain NE point is related to the relative slopes of the best-response functions and is given by [37] [38]:

$$\begin{aligned}
NNM_1 &= 2P_1^2\Gamma_0\left[P_2^2\left(\theta_2^{(q)}\right)^2|h_{21}|^4|h_{1r}|^2+P_2^2\left(\theta_2^{(q)}\right)^2|h_{21}|^2|h_{11}|^2|h_{2r}|^2+\right.\\
&\left.\left(\tilde{N}_1^{(q)}\right)^2|h_{1r}|^2+2\,P_2\,\tilde{N}_1^{(q)}\theta_2^{(q)}\,|h_{21}|^2|h_{1r}|^2\right]\\
&\left(|h_{21}|^2\theta_2^{(q)}\,P_2+\Gamma_0+\tilde{N}_1^{(q)}\right)\left(\theta_2^{(q)}\,P_2\,N_r^{(q)}\,|h_{21}|^2+|h_{2r}|^2\theta_2^{(q)}\,P_2\,\tilde{N}_1^{(q)}+N_r^{(q)}\,\tilde{N}_1^{(q)}\right)\cdot\\
&\left(|h_{1r}h_{21}-h_{11}h_{2r}|^2\theta_2^{(q)}\,P_2+|h_{1r}|^2\tilde{N}_1^{(q)}+|h_{11}|^2N_r^{(q)}\right)\\
&\geq 0
\end{aligned} \tag{31}$$

$$\begin{aligned}
DNM_1 &= -|h_{11}h_{2r}-h_{21}h_{1r}|^2\theta_1^{(q)}P_1\theta_2^{(q)}P_2\tilde{N}_1^{(q)}-|h_{21}|^2|h_{1r}|^2\theta_1^{(q)}P_1\theta_2^{(q)}P_2\tilde{N}_1^{(q)}-|h_{11}|^2\theta_1^{(q)}P_1N_r^{(q)}\tilde{N}_1^{(q)}-\\
&|h_{21}|^2|h_{2r}|^2\left(\theta_2^{(q)}\right)^2P_2^2\tilde{N}_1^{(q)}-2|h_{21}|^2\theta_2^{(q)}P_2N_r^{(q)}\tilde{N}_1^{(q)}-|h_{21}|^4\left(\theta_2^{(q)}\right)^2P_2^2N_r^{(q)}-\\
&|h_{21}h_{1r}-h_{11}h_{2r}|^2|h_{21}|^2\theta_1^{(q)}P_1\left(\theta_2^{(q)}\right)^2P_2^2-|h_{1r}|^2\theta_1^{(q)}P_1\left(\tilde{N}_1^{(q)}\right)^2-|h_{2r}|^2\theta_2^{(q)}P_2\left(\tilde{N}_1^{(q)}\right)^2-\\
&\left(\tilde{N}_1^{(q)}\right)^2N_r^{(q)}-|h_{21}|^2\theta_2^{(q)}P_2N_r^{(q)}\Gamma_0-N_r^{(q)}\Gamma_0\tilde{N}_1^{(q)}-|h_{2r}|^2\theta_2^{(q)}P_2\tilde{N}_1^{(q)}\Gamma_0-\\
&|h_{11}|^2|h_{21}|^2\theta_1^{(q)}P_1\theta_2^{(q)}P_2N_r^{(q)}\\
&\leq 0
\end{aligned}\ . \tag{32}$$

$$\left|\frac{\mathrm{dBR}_1}{\mathrm{d}\theta_2}\frac{\mathrm{dBR}_2}{\mathrm{d}\theta_1}\right|<1 \tag{33}$$

in an open neighborhood of the NE point. We denote by $\mathcal{V}(\theta_1,\theta_2)$ an open neighborhood of $(\theta_1,\theta_2)\in[0,1]^2$.

1) If $d_1\leq 0$ and $d_2\leq 0$, then the BR are constants $\mathrm{BR}_i(\theta_j)=0$ and thus the NE is unique $(\theta_1^{\mathrm{NE}},\theta_2^{\mathrm{NE}})=(0,0)$, for all $c_{ii}\geq 0$, $c_{ji}\geq 0$. The condition (33) is met since $\left|\frac{\mathrm{dBR}_1}{\mathrm{d}\theta_2}\frac{\mathrm{dBR}_2}{\mathrm{d}\theta_1}\right|=0$ for $(\theta_1,\theta_2)\in\mathcal{V}(0,0)$ and thus the NE is asymptotically stable.
2) If $d_1\leq 0$ and $d_2>0$, then it can be checked that the NE is unique, for all $c_{ii}\geq 0$, $c_{ji}\geq 0$: $\theta_1^{\mathrm{NE}}=0$ and

$$\theta_2^{\mathrm{NE}}=\left|\begin{array}{lll}\frac{d_2}{c_{22}} & \text{, if} & d_2<c_{22},\\ 1 & \text{, otherwise.} & \end{array}\right.$$

It can be checked that $\left|\frac{\mathrm{dBR}_1}{\mathrm{d}\theta_2}\frac{\mathrm{dBR}_2}{\mathrm{d}\theta_1}\right|=0$ for $(\theta_1,\theta_2)\in\mathcal{V}(\theta_1^{\mathrm{NE}},\theta_2^{\mathrm{NE}})$ and the NE is asymptotically stable.
3) If $d_1>0$ and $d_2\leq 0$, then, similarly to the previous item, we have a unique NE, for all $c_{ii}\geq 0$,

$c_{ji} \geq 0$: $\theta_2^{\mathrm{NE}} = 0$ and

$$\theta_1^{\mathrm{NE}} = \begin{vmatrix} \frac{d_1}{c_{11}} & \text{, if} & d_1 < c_{11}, \\ 1 & \text{, otherwise.} & \end{vmatrix}$$

Here as well we have $\left|\frac{\mathrm{dBR}_1}{\mathrm{d}\theta_2}\frac{\mathrm{dBR}_2}{\mathrm{d}\theta_1}\right| = 0$ for $(\theta_1, \theta_2) \in \mathcal{V}(\theta_1^{\mathrm{NE}}, \theta_2^{\mathrm{NE}})$ and the NE is asymptotically stable.

4) If $d_1 > 0$ and $d_2 > 0$, we have to take into consideration the parameters $c_{ii} \geq 0$, $c_{ji} \geq 0$.
   a) If $F_1(1) \geq 1$ and $F_2(1) \geq 1$, then we have $d_1 \geq c_{12} + c_{11}$ and $d_2 \geq c_{21} + c_{22}$. In this case the BR are constants i.e., $\mathrm{BR}_i(\theta_j) = 1$ and thus the NE is unique $(\theta_1^{\mathrm{NE}}, \theta_2^{\mathrm{NE}}) = (1, 1)$. We have $\left|\frac{\mathrm{dBR}_1}{\mathrm{d}\theta_2}\frac{\mathrm{dBR}_2}{\mathrm{d}\theta_1}\right| = 0$ for $(\theta_1, \theta_2) \in \mathcal{V}(1, 1)$ and the NE is asymptotically stable.
   b) If $F_1(1) \geq 1$ and $F_2(1) < 1$, then we have $d_1 \geq c_{12} + c_{11}$ and $d_2 < c_{21} + c_{22}$. Here also the NE is unique and $\theta_1^{\mathrm{NE}} = 1$ and

$$\theta_2^{\mathrm{NE}} = \begin{vmatrix} \frac{d_2 - c_{21}}{\alpha_{22}} & \text{, if} & d_2 > c_{22}, \\ 0 & \text{, otherwise.} & \end{vmatrix}$$

   Similarly, we have $\left|\frac{\mathrm{dBR}_1}{\mathrm{d}\theta_2}\frac{\mathrm{dBR}_2}{\mathrm{d}\theta_1}\right| = 0$ for $(\theta_1, \theta_2) \in \mathcal{V}(\theta_1^{\mathrm{NE}}, \theta_2^{\mathrm{NE}})$ and the NE is asymptotically stable.
   c) If $F_1(1) < 1$ and $F_2(1) \geq 1$, then we have $d_1 < c_{12} + c_{11}$ and $d_2 \geq c_{21} + c_{22}$. Here also the NE is unique and $\theta_2^{\mathrm{NE}} = 1$ and

$$\theta_1^{\mathrm{NE}} = \begin{vmatrix} \frac{d_1 - c_{12}}{c_{11}} & \text{, if} & d_1 > c_{11}, \\ 0 & \text{, otherwise.} & \end{vmatrix}$$

   Here as well we have $\left|\frac{\mathrm{dBR}_1}{\mathrm{d}\theta_2}\frac{\mathrm{dBR}_2}{\mathrm{d}\theta_1}\right| = 0$ for $(\theta_1, \theta_2) \in \mathcal{V}(\theta_1^{\mathrm{NE}}, \theta_2^{\mathrm{NE}})$ and the NE is asymptotically stable.
   d) If $F_1(1) < 1$ and $F_2(1) < 1$, then we'll have $d_1 < c_{12} + c_{11}$ and $d_2 < c_{21} + c_{22}$. This case is the most demanding one and will be treated in detail separately.

At this point an important observation is in order. The discussed scenarios, for which we have determined the unique NE, have a simple geometric interpretation. If the intersection point $(\theta_1^*, \theta_2^*)$ is such that either $\theta_1^* \in \mathbb{R} \setminus [0, 1]$ or $\theta_2^* \in \mathbb{R} \setminus [0, 1]$ then the NE is unique and differs from this point $((\theta_1^{\mathrm{NE}}, \theta_2^{\mathrm{NE}}) \neq (\theta_1^*, \theta_2^*))$. The case 4.(d) corresponds to the case where the intersection point $(\theta_1^*, \theta_2^*) \in [0, 1]^2$ is an NE point. Now we are interested in finding whether this intersection point is the unique NE or there are more than one NE. If $0 < d_1 < c_{11} + c_{12}$ and $0 < d_2 < c_{22} + c_{21}$ we have the following situations:

1) If $c_{11}c_{22}=c_{21}c_{12}$, then the curves described by $\theta_i=F_i(\theta_j)$ are parallel.

   a) If $d_1=d_2$, then the curves are superposed. In this special case we have an infinity of NE that can be characterized by $(\theta_1^{\mathrm{NE}},\theta_2^{\mathrm{NE}})\in\mathcal{T}$ where:

$$\mathcal{T}=\left\{(\theta_1,\theta_2)\in[0,1]^2\left|\theta_1=F_1(\theta_2^{\mathrm{NE}})\right.\right\}.$$

   In this case we have an infinity of NE such that $\left|\frac{\mathrm{dBR}_1}{\mathrm{d}\theta_2}\frac{\mathrm{dBR}_2}{\mathrm{d}\theta_1}\right|=1$ for $(\theta_1,\theta_2)\in\mathcal{V}(\theta_1^{\mathrm{NE}},\theta_2^{\mathrm{NE}})$ and the NEs are not stable states. This can be easily understood since a small deviation from a certain NE drives the users to a new NE point. Thus, the users don't return to the initial state.

   b) If $d_1\neq d_2$, then the two lines are only parallel. In this case it can be checked that the NE is unique and also asymptotically stable since again $\left|\frac{\mathrm{dBR}_1}{\mathrm{d}\theta_2}\frac{\mathrm{dBR}_2}{\mathrm{d}\theta_1}\right|=0$ for $(\theta_1,\theta_2)\in\mathcal{V}(\theta_1^{\mathrm{NE}},\theta_2^{\mathrm{NE}})$. In order to explicit the exact relation of the NE, one has to consider all scenarios in function of the sign of the following four relations $F_i(0)-1$ and $\theta_j^o-1$, $i\in\{1,2\}$. We will explicit only one of them. Let us assume that $F_i(0)-1<0$ and $\theta_j^o<0$ which means that $d_1<\min\{c_{12},\frac{c_{12}c_{21}}{c_{22}}\}$ and $d_2<\min\{c_{21},c_{22}\}$. Here we have two sub-cases:

   - If $\frac{d_1}{c_{12}}<\frac{d_2}{c_{22}}$, then the NE is characterized by $\theta_1^{\mathrm{NE}}=0$ and $\theta_2^{\mathrm{NE}}=\frac{d_2}{c_{22}}$.
   - If $\frac{d_1}{c_{12}}>\frac{d_2}{c_{22}}$, then the NE is characterized by $\theta_1^{\mathrm{NE}}=\frac{d_1c_{22}}{c_{12}c_{21}}$ and $\theta_2^{\mathrm{NE}}=0$.

2) Consider $c_{11}c_{22}\neq c_{21}c_{12}$. Here we have to consider all cases in function of the sign of the four relations $F_i(0)-1$ and $\theta_j^o-1$, $i\in\{1,2\}$. We will focus on only one of them. Let us assume that $F_i(0)-1<0$ and $\theta_j^o-1<0$ and thus $d_1<\min\{c_{12},c_{11}\}$ and $d_2<\min\{c_{21},c_{22}\}$. Here we have four sub-cases:

   - If $\frac{d_2}{c_{22}}<\frac{d_1}{c_{12}}$ and $\frac{d_1}{c_{11}}>\frac{d_2}{c_{21}}$, then the NE is unique: $\theta_1^{\mathrm{NE}}=\theta_1^*$ and $\theta_2^{\mathrm{NE}}=\theta_2^*$. Also we have that $\left|\frac{\mathrm{dBR}_1}{\mathrm{d}\theta_2}\frac{\mathrm{dBR}_2}{\mathrm{d}\theta_1}\right|<1$ for $(\theta_1,\theta_2)\in\mathcal{V}(\theta_1^*,\theta_2^*)$ and the NE is asymptotically stable.
   - If $\frac{d_2}{c_{22}}>\frac{d_1}{c_{12}}$ and $\frac{d_1}{c_{11}}<\frac{d_2}{c_{21}}$, then there are three different NE: $(\theta_1^{\mathrm{NE}},\theta_2^{\mathrm{NE}})\in\{(\theta_1^*,\theta_2^*),(0,\frac{d_2}{c_{22}}),(\frac{d_1}{c_{11}},0)\}$, the intersection point and two other NE's on the border. The intersection point is unstable since $\left|\frac{\mathrm{dBR}_1}{\mathrm{d}\theta_2}\frac{\mathrm{dBR}_2}{\mathrm{d}\theta_1}\right|>1$ for $(\theta_1,\theta_2)\in\mathcal{V}(\theta_1^*,\theta_2^*)$ but the other two NE's are asymptotically stable since $\left|\frac{\mathrm{dBR}_1}{\mathrm{d}\theta_2}\frac{\mathrm{dBR}_2}{\mathrm{d}\theta_1}\right|=0$ for $(\theta_1,\theta_2)\in\mathcal{V}(0,\frac{d_2}{c_{22}})$ and $(\theta_1,\theta_2)\in\mathcal{V}(\frac{d_1}{c_{11}},0)$.
   - If $\frac{d_2}{c_{22}}=\frac{d_1}{c_{12}}$ and $\frac{d_1}{c_{11}}<\frac{d_2}{c_{21}}$, then there are only two different NE: $(\theta_1^{\mathrm{NE}},\theta_2^{\mathrm{NE}})\in\{(0,\frac{d_2}{c_{22}}),(\frac{d_1}{c_{11}},0)\}$. In this case both of NEs are on the border, one of which represents the intersection point of the BR's. It turns out that the intersection point is not a stable NE because $\left|\frac{\mathrm{dBR}_1}{\mathrm{d}\theta_2}\frac{\mathrm{dBR}_2}{\mathrm{d}\theta_1}\right|>1$ for $(\theta_1, \theta_2) \in \mathcal{V}(0, \frac{d_2}{c_{22}})$. However, the other NE is asymptotically stable since $\left|\frac{\mathrm{dBR}_1}{\mathrm{d}\theta_2}\frac{\mathrm{dBR}_2}{\mathrm{d}\theta_1}\right| = 0$ for $(\theta_1, \theta_2) \in \mathcal{V}(\frac{d_1}{c_{11}}, 0)$.

- If $\frac{d_2}{c_{22}} > \frac{d_1}{c_{12}}$ and $\frac{d_1}{c_{11}} = \frac{d_2}{c_{21}}$, then there are two NE: $(\theta_1^{\mathrm{NE}}, \theta_2^{\mathrm{NE}}) \in \{(\frac{d_1}{c_{11}}, 0), (0, \frac{d_2}{c_{22}})\}$. Here the analysis of the stability of the two NE's is similar to the previous case.

In conclusion, the number of NE states depends on the geometrical properties of the best-response functions. Three different cases can be identified: 1) when the lines $\theta_i = F_i(\theta_j)$ are superposed the game has an infinity of NE which are not stable; 2) when the lines have a unique intersection point that lies outside of the borders $[0,1] \times [0,1]$, the NE is unique and asymptotically stable; 3) when the lines have a unique intersection point $(\theta_1^*, \theta_2^*)$ that lies inside $[0,1] \times [0,1]$, there can be one, two or three different NE among which one is identical to this intersection point. In the case where the the NE is unique, it is also asymptotically stable. When the game has two or three NE, the intersection point $(\theta_1^*, \theta_2^*)$ is an unstable equilibrium while the other/others are asymptotically stable. The best-response algorithm converges to one of the NE points depending on the initial state of the system.

## References

[1] E. V. Belmega, B. Djeumou, and S. Lasaulce, "What happens when cognitive terminals compete for a relay node?" in *IEEE International Conference on Accoustics, Speech and Signal Processing (ICASSP)*, Taipei, Taiwan, Apr. 2009, pp. 2609–2612.

[2] ——, "Resource allocation games in interference relay channels," in *IEEE International Conference on Game Theory for Networks (GAMENETS)*, Istanbul, Turkey, May 2009, pp. 575–584.

[3] O. Sahin and E. Erkip, "Achievable rates for the Gaussian interference relay channel," in *Proc. IEEE Global Communications Conference (GLOBECOM'07)*, Washington D.C., USA, Nov. 2007, pp. 786–787.

[4] ——, "On achievable rates for interference relay channel with interference cancellation," in *Proc. IEEE Annual Asilomar Conference on Signals, Systems and Computers (invited paper)*, Pacific Grove, CA, USA, Nov. 2007, pp. 805–809.

[5] W. Yu, W. Ree, S. Boyd, and J. M. Cioffi, "Iterative water-filling for Gaussian vector multiple-access channels," *IEEE Trans. Inf. Theory*, vol. 50, no. 1, pp. 145–152, Jan. 2004.

[6] S. T. Chung, S. J. Kim, J. Lee, and J. M. Cioffi, "A game theoretic approach to power allocation in frequency-selective gaussian interference channels," in *Proc. IEEE Intl. Symposium on Information Theory (ISIT)*, Pacifico Yokohama, Kanagawa, Japan, Jun./Jul. 2003, pp. 316–316.

[7] Z. Q. Luo and J. S. Pang, "Analysis of iterative waterfilling algorithm for multiuser power control in digital subscriber lines," *Eurasip Journal on Applied Signal Processing*, pp. 1–10, 2006.

[8] M. Bennis, M. L. Treust, S. Lasaulce, and M. Debbah, "Spectrum sharing games on the interference channel," in *Proc. IEEE Intl. Conf. on Game Theory for Networks (Gamenets)*, Istanbul, Turkey, May 2010.

[9] Y. Xi and E. M. Yeh, "Equilibria and price of anarchy in parallel relay networks with node pricing," in *Conference on Information Sciences and Systems (CISS)*, Princeton, NJ, USA, Mar. 2008, pp. 944–949.

[10] ——, "Pricing, competition, and routing for selfish and strategic nodes in multi-hop relay networks," in *IEEE Conference on Computer Communications (INFOCOM)*, Phoneix, AZ, USA, Apr. 2008, pp. 1463–1471.

[11] Y. Shi, Y. Wang, W. L. Huang, and K. B. Letaief, “Power allocation in Gaussian interference relay channel via game theory,” in *IEEE Global Telecommunications Conference (GLOBECOM)*, New Orleans, LO, USA, Dec. 2008, pp. 944–949.

[12] C. H. Papadimitriou, “Algorithms, games, and the internet,” in *Annual ACM Symposium on Theory of Computing (STOC)*, Heraklion, Crete, Greece, Jul. 2001, pp. 749–753.

[13] I. Maric, R. Dabora, and A. Goldsmith, “On the capacity of the interference channel with a relay,” in *IEEE International Symposium on Information Theory*, Toronto, Canada, Jul. 2008, pp. 554–558.

[14] T. M. Cover and A. A. El Gamal, “Capacity theorems for the relay channel,” *IEEE Trans. Inf. Theory*, vol. 25, no. 5, pp. 572–584, Sep. 1979.

[15] I. Maric, R. Dabora, and A. Goldsmith, “Generalized relaying in the presence of interference,” in *The 42nd Asilomar Conference on Signals, Systems and Computers*, Pacific Grove, CA, USA, Oct. 2008.

[16] R. Dabora, I. Maric, and A. Goldsmith, “Interference forwarding in multiuser networks,” in *IEEE Global Telecommunications Conference (GLOBECOM)*, New Orleans, LA, USA, Nov. 2008, pp. 1–5.

[17] ——, “Relay strategies for interference-forwarding,” in *IEEE Information Theory Workshop (ITW)*, Porto, Portugal, May 2008.

[18] A. D. Wyner and J. Ziv, “The rate-distortion function for source coding with side information at the decoder,” *IEEE Trans. Inf. Theory*, vol. IT-22, no. 1, pp. 1–11, Jan. 1976.

[19] D. Gunduz, E. Tuncel, and J. Nayak, “Rate regions for the separated two-way relay channel,” in *The 46th Allerton Conference on Communication, Control, and Computing*, Monticello, IL, USA, Sep. 2008.

[20] T. M. Cover and J. A. Thomas, “Elements of information theory,” *Wiley Interscience*, 2006.

[21] I. Abou-Faycal and M. Médard, “Optimal uncoded regeneration for binary antipodal signaling,” in *Proc. IEEE Int. Conf. on Communications (ICC)*, Paris, France, Jun. 2004, pp. 742–746.

[22] M. A. Khojastepour, A. Sabharwal, and B. Aazhang, “Lower bounds on the capacity of Gaussian relay channel,” in *Proc. IEEE Conf. on Information Sciences and Systems (CISS)*, Princeton, NJ, USA, Mar. 2004, pp. 597–602.

[23] K. S. Gomadam and S. A. Jafar, “On the capacity of memoryless relay networks,” in *Proc. IEEE Int. Conf. on Communications (ICC)*, Istanbul, Turkey, Jun. 2006, pp. 1580–1585.

[24] B. Djeumou, S. Lasaulce, and A. G. Klein, “Practical quantize-and-forward schemes for the frequency division relay channel,” *EURASIP Journal on Wireless Communications and Networking*, vol. 2007, Article ID 20258, 11 pages, doi:10.1155/2007/20258, 2007.

[25] A. A. El Gamal, M. Mohseni, and S. Zahedi, “Bounds on capacity and minimum energy-per-bit for AWGN relay channels,” *IEEE Trans. Inf. Theory*, vol. 52, no. 4, pp. 1545–1561, Apr. 2006.

[26] J. F. Nash, “Equilibrium points in n-points games,” *Proc. of the Nat. Academy of Science*, vol. 36, no. 1, pp. 48–49, Jan. 1950.

[27] J. Rosen, “Existence and uniqueness of equilibrium points for concave n-person games,” *Econometrica*, vol. 33, pp. 520–534, 1965.

[28] S. Boyd and L. Vandenberghe, “Convex optimization,” *Cambridge University Press*, 2004.

[29] D. Monderer and L. S. Shapley, “Potential games,” *Econometrica*, vol. 14, pp. 124–143, 1996.

[30] D. M. Topkis, “Supermodularity and complementarity,” *Princeton University Press*, 1998.

[31] R. Mochaourab and E. Jorswieck, “Resource allocation in protected and shared bands: Uniqueness and efficiency of Nash equilibria,” in *Fourth International Conference on Performance Evaluation Methodologies and Tools (Valuetools)*, Pisa, Italy, Oct. 2009.

[32] E. Jorswieck and R. Mochaourab, “Power control game in protected and shared bands: Manipulability of Nash equilibrium,” in *IEEE International Conference on Game Theory for Network (GAMENETS)*, Istanbul, Turkey, May 2009, pp. 428–437.

[33] A. Cournot, *Recherches sur les principes mathématiques de la la théorie des richesses*, (Re-edited by Mac Millan in 1987) 1838.

[34] H. Moulin, “Dominance solvability and Cournot stability,” *Mathematical Social Sciences*, vol. 7, pp. 83–102, 1984.

[35] H. von Stackelberg, *The theory of the market economy*. Oxford, England: Oxford University Press, 1952.

[36] A. B. Carleial, "Interference channels," *IEEE Trans. Inf. Theory*, vol. 24, no. 1, pp. 60–70, Jan. 1978.

[37] D. Fudenberg and J. Tirole, "Game theory," *The MIT Press*, 1991.

[38] H. R. Varian, "Microeconomic analysis," *W. W. Norton and Company*, 1992.

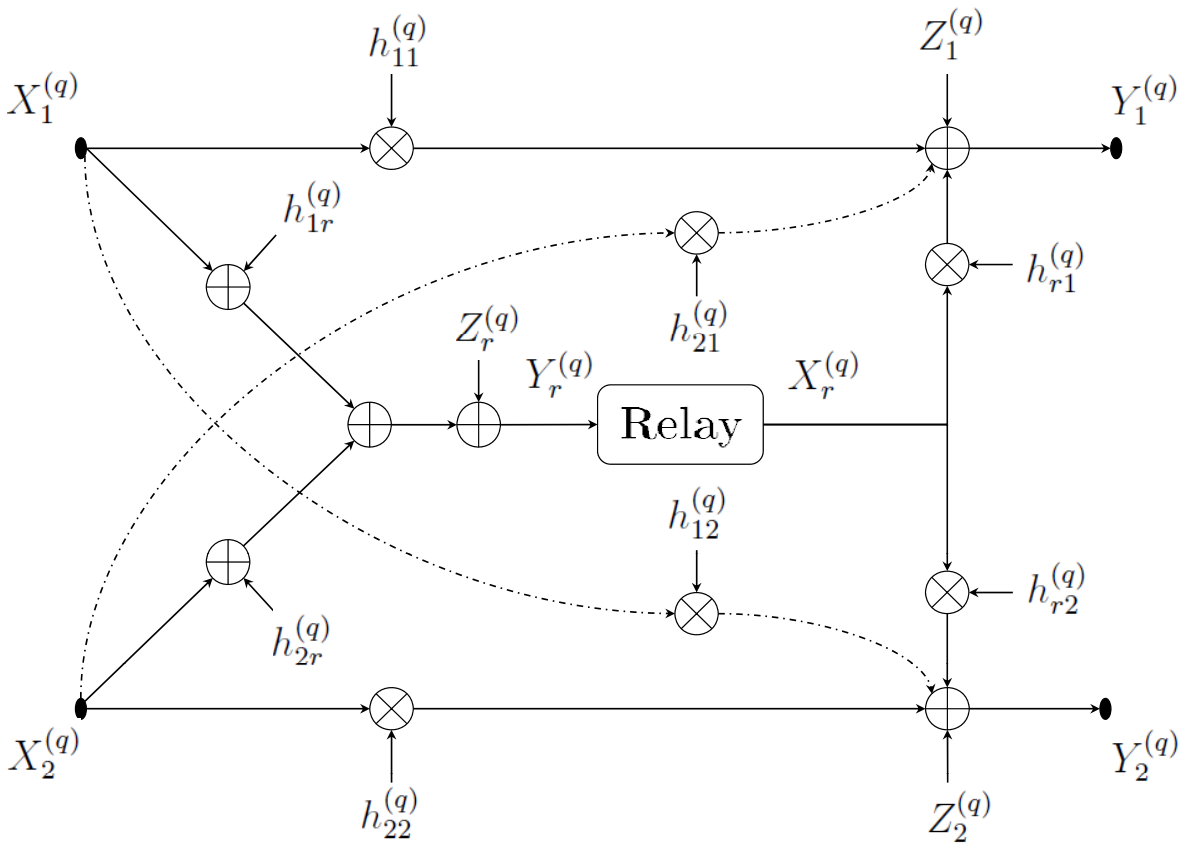

Fig. 1. System model: a $Q$-band interference channel with a relay; $q$ is the band index and $q \in \{1, ..., Q\}$.

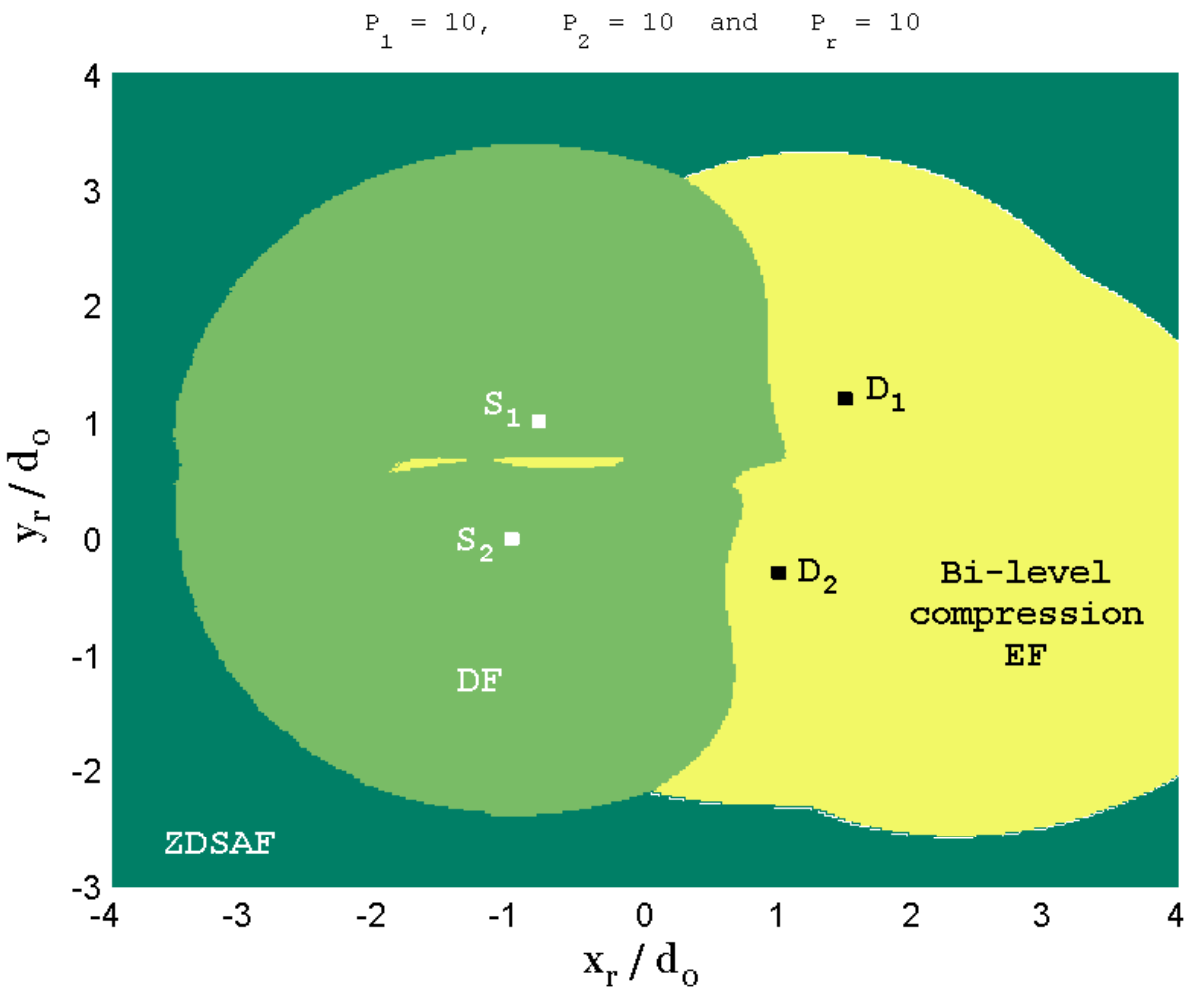

Fig. 2. For different relay positions in the plane $\left(\frac{x_r}{d_0}, \frac{y_r}{d_0}\right) \in [-4, +4] \times [-3, +4]$, the figure indicates the regions where one relaying protocol (AF, DF or bi-level EF) dominates the two others in terms of network sum-rate.

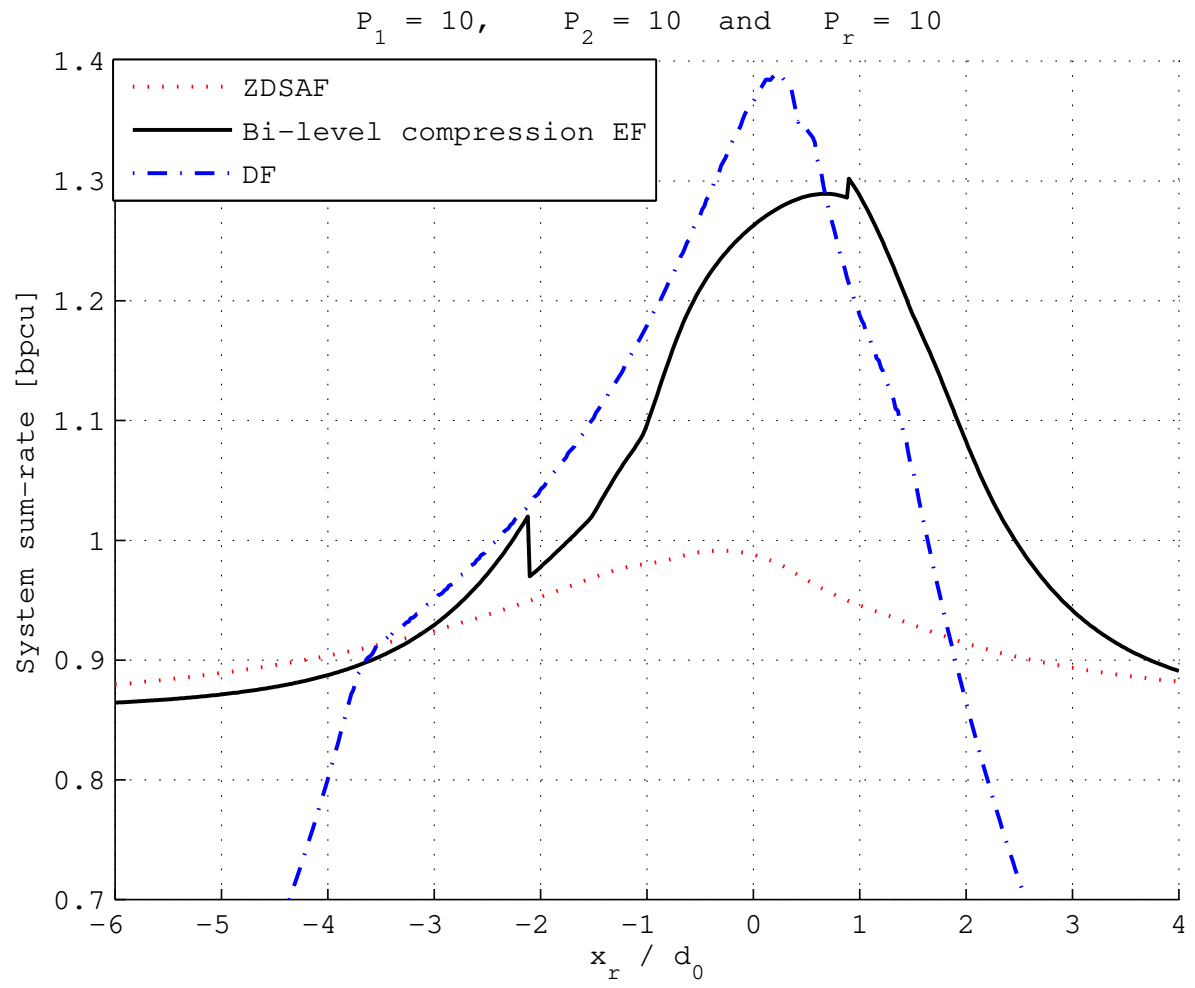

Fig. 3. Achievable system sum-rate versus $x_r$ (abscissa for the relay position) for a fixed $y_r$ ($y_r = 0.5d_0$), with AF, DF and bi-level EF.

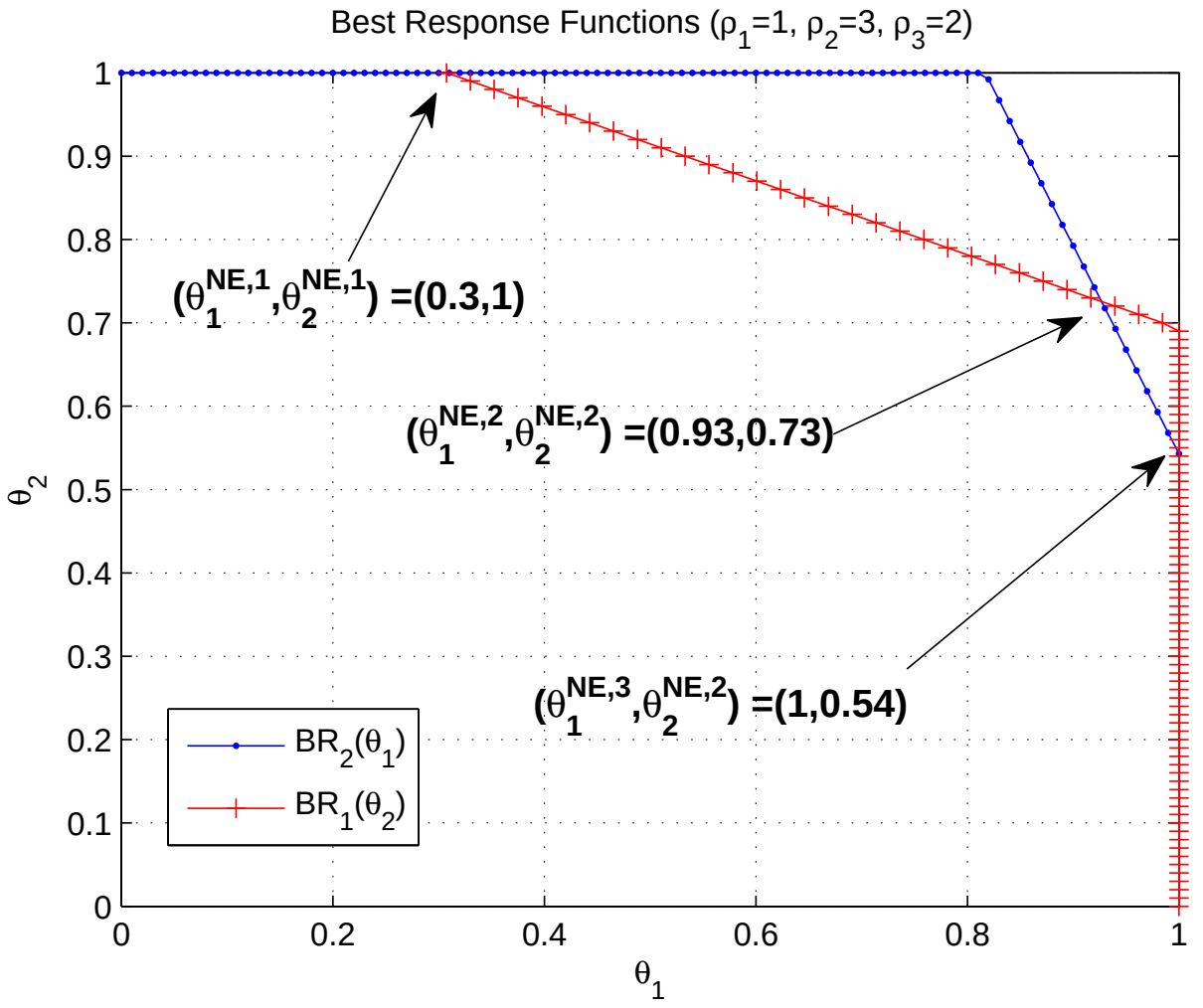

Fig. 4. Best replies for a system composed of an IC in band (1) and IRC in band (2) when the ZDSAF protocol is assumed (fixed amplification factor). The number of equilibria is generally three as indicated the figure.

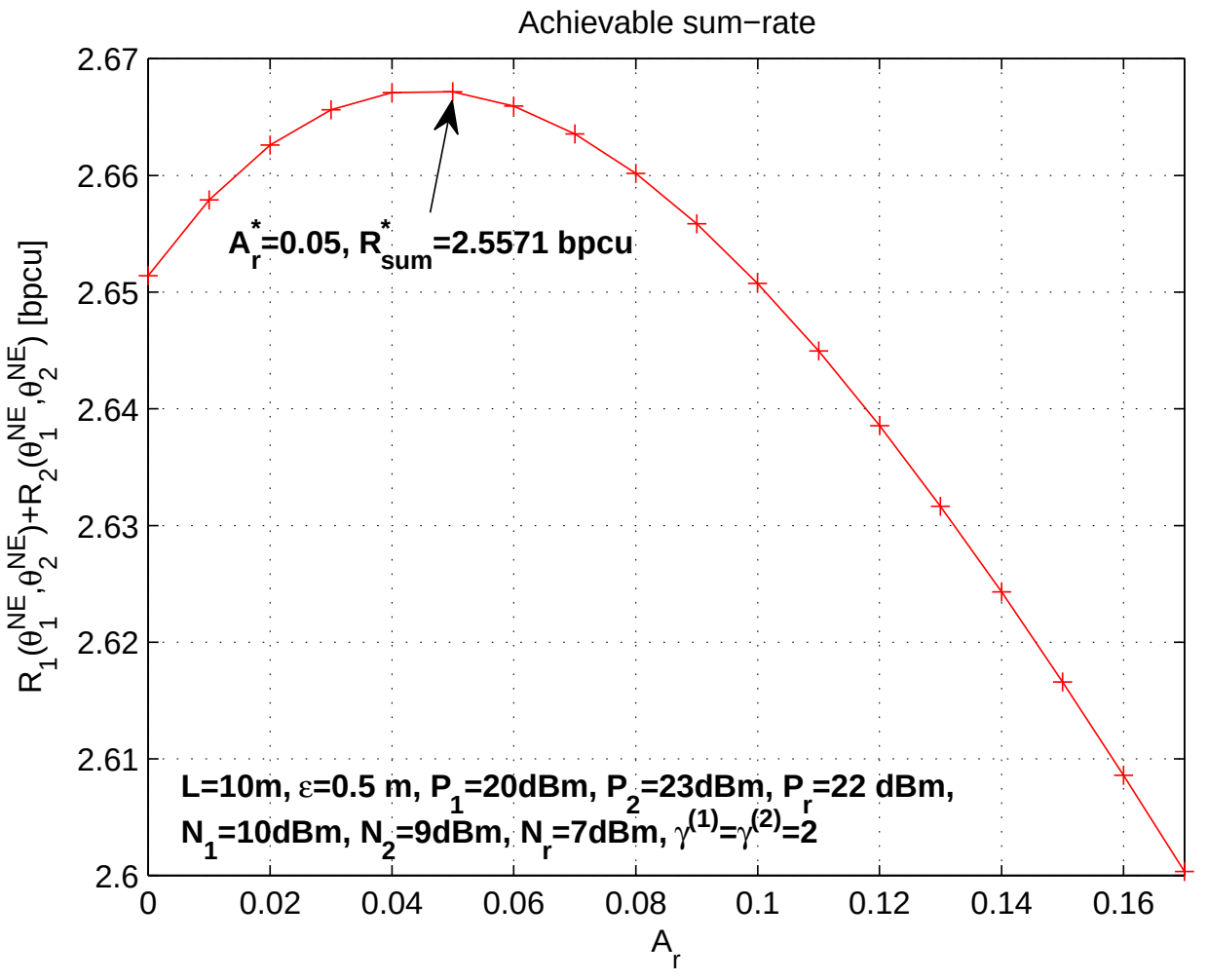

Fig. 5. ZDSAF relaying protocol with fixed amplification gain. Achievable network sum-rate at the NE as a function of $A_r \in [0, \tilde{a}_r]$ for $L = 10\text{m}$, $\epsilon = 0.5\text{m}$, $P_1 = 20\text{dBm}$, $P_2 = 23\text{dBm}$, $P_r = 22\text{dBm}$, $N_1 = 10\text{dBm}$, $N_2 = 9\text{dBm}$, $N_r = 7\text{dBm}$, $\gamma^{(1)} = \gamma^{(2)} = 2$. The optimal amplification gain $A_r^* = 0.05 \leq \tilde{a}_r(1,1) = 0.17$ meaning that saturating the relay power constraint is suboptimal.

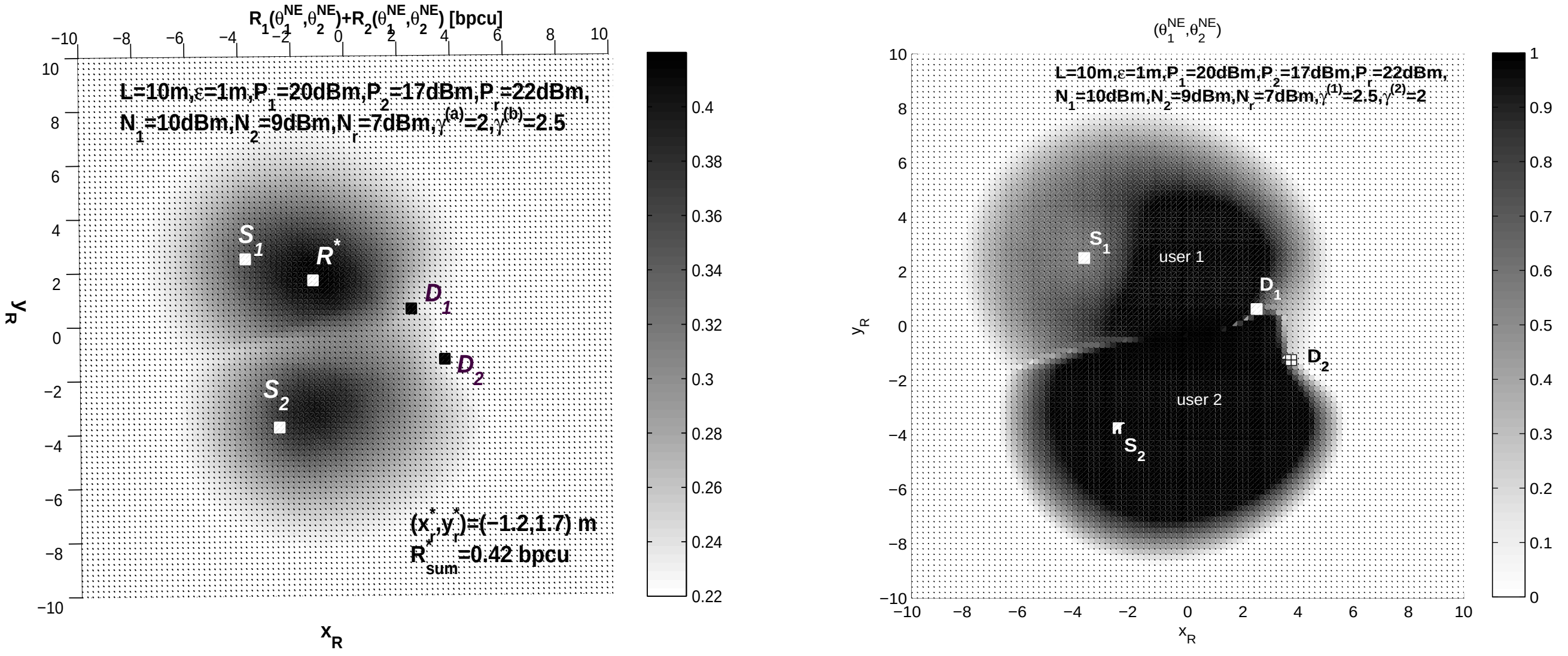

(a) (b)

Fig. 6. ZDSAF relaying protocol, full power regime. $L = 10\text{m}$, $\epsilon = 1\text{m}$, $P_1 = 20\text{dBm}$, $P_2 = 17\text{dBm}$, $P_r = 22\text{dBm}$, $N_1 = 10\text{dBm}$, $N_2 = 9\text{dBm}$, $N_r = 7\text{dBm}$, $\gamma^{(1)} = 2.5$ and $\gamma^{(2)} = 2$. (a) Achievable network sum-rate at the NE as a function of $(x_{\mathcal{R}}, y_{\mathcal{R}}) \in [-L, L]^2$ (the optimal relay position $(x_{\mathcal{R}}^*, y_{\mathcal{R}}^*) = (-1.2, 1.7)$ lies on the segment between $\mathcal{S}_1$ and $\mathcal{D}_1$). (b) Power allocation policies at the NE $(\theta_1^{NE}, \theta_2^{NE})$ as a function of $(x_{\mathcal{R}}, y_{\mathcal{R}}) \in [-L, L]^2$ (the regions where the uses allocate their power to IRC are almost non overlapping.).

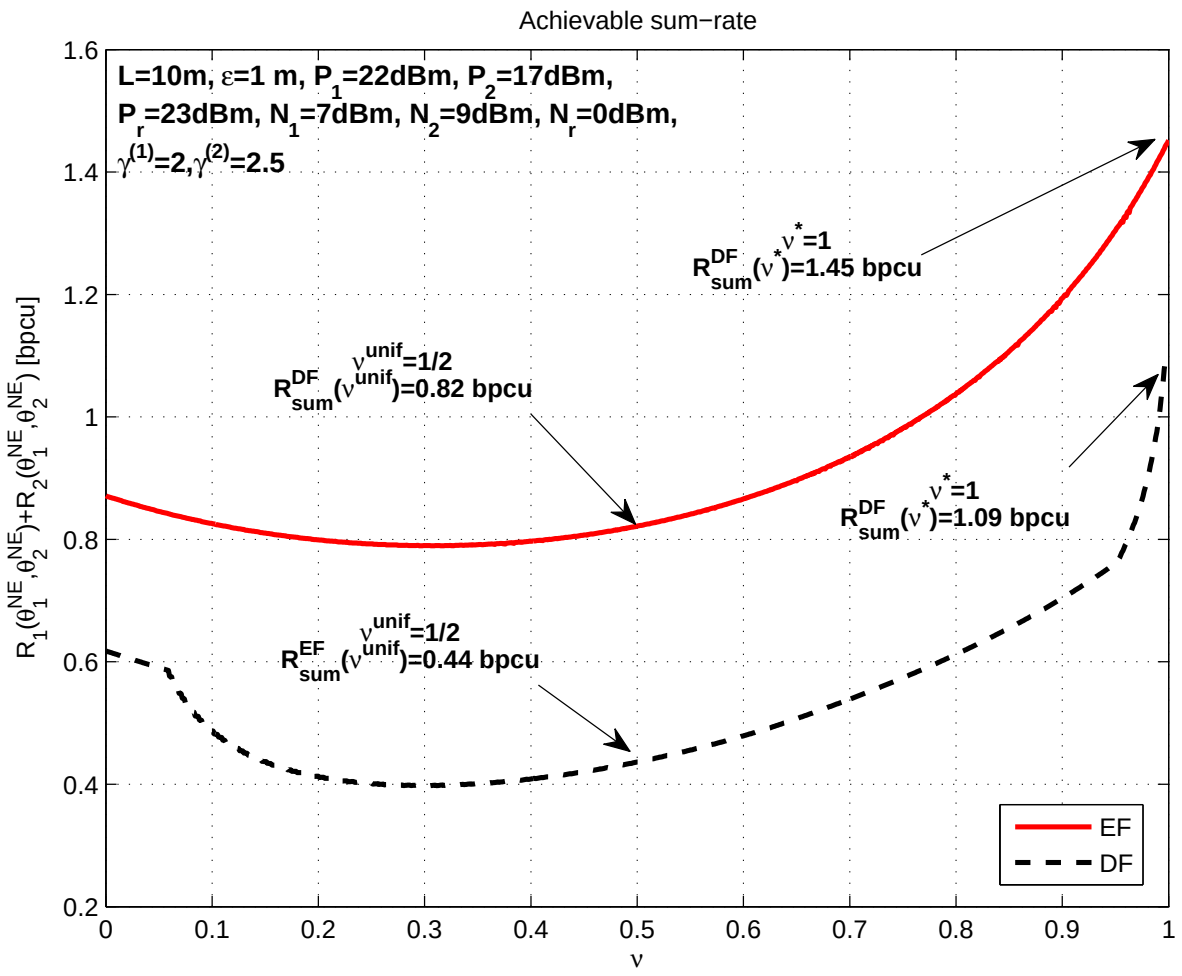

Fig. 7. EF vs. DF relaying protocol. Achievable network sum-rate at the NE as a function of $\nu \in [0, 1]$ for $L = 10\text{m}$, $\epsilon = 1\text{m}$, $P_1 = 22\text{dBm}$, $P_2 = 17\text{dBm}$, $P_r = 23\text{dBm}$, $N_1 = 7\text{dBm}$, $N_2 = 9\text{dBm}$, $N_r = 0\text{dBm}$, $\gamma^{(1)} = 2.5$ and $\gamma^{(2)} = 2$. The optimal relay PA $\nu^* = 1$ is in favor of the better user and outperforms the uniform relay PA $\nu = 0.5$ for both EF and DF.